\documentclass[useAMS,usenatbib]{./mnras}

\usepackage{times,mathptm}
\usepackage{lscape}
\usepackage{natbib}
\usepackage{graphicx}
\usepackage{./floatflt}
\usepackage{amssymb}
\newcommand{\comment}[1]{}
\newcommand{\e}[1]{$\times 10^{#1}$}

\newcommand{\mum}{${\rm \umu m}$}
\newcommand{\mujy}{$\umu$Jy}
\newcommand{\ergs}{${\rm ergs~s^{-1}}$}
\newcommand{\kev}{keV}
\newcommand{\msun}{M$_\odot$}
\newcommand{\lsun}{L$_\odot$}

\newcommand{\sig}{$\sigma$}

\newcommand{\nh}{$N_{\rm H}$}
\newcommand{\lx}{$L_{\rm X}$}
\newcommand{\Lx}{$L_{\rm X}$}
\newcommand{\lir}{$L_{\rm IR}$}

\newcommand{\mbmb}{$M_{\rm BH}$--$M_{\rm Bulge}$}
\newcommand{\z}{$z$}

\newcommand{\mgal}{$M_{\rm stars}$}

\newcommand{\fig}[1]{fig. \ref{#1}}
\newcommand{\tab}[1]{table \ref{#1}}

\newcommand{\Fig}[1]{Fig. \ref{#1}}

\newcommand{\Spitzer}{{\it Spitzer}}
\newcommand{\IRS}{{\it IRS}}
\newcommand{\Herschel}{{\it Herschel}}

\newcommand{\Chandra}{{\it Chandra}}
\newcommand{\Swift}{{\it Swift}}
\newcommand{\IRAS}{{\it IRAS}}

\title[GOODS-H: Star formation in AGN host galaxies] {GOODS-{\it
    Herschel}: The far-infrared view of star formation in AGN host
  galaxies since $z\approx3$}

\author[J. R. Mullaney et al.]
{J.~R.~Mullaney$^{1}$\thanks{E-mail: james.mullaney@cea.fr}, 
M.~Pannella$^{1}$,
E.~Daddi$^{1}$,
D.~M.~Alexander$^{2}$,
D.~Elbaz$^{1}$,
R.~C.~Hickox$^{2}$,
\newauthor
F.~Bournaud$^{1}$, 
B.~Altieri$^{3}$,
H.~Aussel$^{1}$,
D.~Coia$^{3}$,
H.~Dannerbauer$^{1}$,
K.~Dasyra$^{1,4}$,
\newauthor
M.~Dickinson$^{5}$,
H.~S.~Hwang$^{1}$,
J.~Kartaltepe$^{5}$,
R.~Leiton$^{1}$,
G.~Magdis$^{1}$,
B.~Magnelli$^{6}$,
\newauthor
P.~Popesso$^{6}$,
I.~Valtchanov$^{3}$,
F.~E.~Bauer$^{7}$,
W.~N.~Brandt$^{8}$,
A.~Del~Moro$^{2}$, 
D.~J.~Hanish$^{9}$,
\newauthor
R.~J.~Ivison$^{10}$, 
S.~Juneau$^{11}$,
B.~Luo$^{12}$, 
D.~Lutz$^{6}$,
M.T.~Sargent$^{1}$, 
D.~Scott$^{13}$,
Y.~Q.~Xue$^{8}$
\\
  $^{1}$Laboratoire AIM, CEA/DSM-CNRS-Universit\'{e} Paris Diderot,
  Irfu/Service d’Astrophysique, CEA-Saclay, Orme des Merisiers,\\
  \ \ 91191 Gif-sur-Yvette Cedex, France\\
  $^{2}$Department of Physics, Durham University, South Road, Durham,
  DH1 3LE, U.K.\\
  $^{3}$Herschel Science Centre, European Space Astronomy Centre, Villanueva de la Ca\~nada, 28691 Madrid, Spain\\
  $^{4}$Observatoire de Paris, LERMA (CNRS:UMR8112), 61 Av. de l' Observatoire, F-75014, Paris, France\\
  $^{5}$National Optical Astronomy Observatory, 950 North Cherry Avenue, Tucson, AZ 85719, USA\\
  $^{6}$Max-Planck-Institut f\"ur Extraterrestrische Physik (MPE), Postfach 1312, 85741, Garching, Germany\\
  $^{7}$Pontificia Universidad Cat{\'o}lica de Chile, Departamento de Astronom{\'{\i}}a y Astrof{\'{\i}}sica, Casilla 306, Santiago 22, Chile\\
  $^{8}$Department of Astronomy and Astrophysics, Pennsylvania State University, University Park, PA 16802, USA\\
  $^{9}$Infrared Processing and Analysis Center, California Institute of Technology, 100-22, Pasadena, CA 91125, USA\\
  $^{10}$UK Astronomy Technology Centre, Royal Observatory, Blackford Hill, Edinburgh EH9 3HJ, UK\\
  $^{11}$Steward Observatory, University of Arizona, Tucson, AZ 85721, USA\\
  $^{12}$Harvard-Smithsonian Center for Astrophysics, 60 Garden Street, Cambridge, MA 02138, USA\\
  $^{13}$Department of Physics and Astronomy, University of British Columbia, Vancouver, BC V6T 1Z1, Canada}

\begin{document}

\date{Date Accepted}

\pagerange{\pageref{firstpage}--\pageref{lastpage}} \pubyear{2007}

\maketitle

\label{firstpage}

\begin{abstract}
  We present a study of the infrared properties of X-ray selected,
  moderate luminosity (i.e, \lx$=10^{42}-10^{44}$~\ergs) active
  galactic nuclei (AGNs) up to z$\approx$3, in order to explore the
  links between star formation in galaxies and accretion onto their
  central black holes. We use 100~\mum\ and 160~\mum\ fluxes from
  GOODS-\Herschel\ -- the deepest survey yet undertaken by the
  \Herschel\ telescope -- and show that in the vast majority of cases
  (i.e., $>$ 94 per cent) these fluxes are dominated by emission from
  the host galaxy.  As such, these far-infrared bands provide an
  uncontaminated view of star formation in the AGN host galaxies.  We
  find no evidence of any correlation between the X-ray and infrared
  luminosities of moderate AGNs at any redshift, suggesting that
  star-formation is decoupled from nuclear (i.e., AGN) activity in
  these galaxies.  On the other hand, we confirm that the star
  formation rates of AGN hosts increase strongly with redshift; by a
  factor of $43^{+27}_{-18}$
 from $z<0.1$ to $z=2-3$
  for AGNs {\it with the same range of X-ray luminosities}.  This
  increase is entirely consistent with the factor of 25--50 increase
  in the specific star formation rates (SSFRs) of normal, star-forming
  (i.e., main-sequence) galaxies over the same redshift range.
  Indeed, the average SSFRs of AGN hosts are only marginally (i.e.,
  $\approx20$ per cent) lower than those of main-sequence galaxies at
  all surveyed redshifts, with this small deficit being due to a
  fraction of AGNs residing in quiescent (i.e., low-SSFR) galaxies.
  We estimate that $79\pm10$ per cent of moderate luminosity AGNs are
  hosted in main-sequence galaxies, $15\pm7$ per cent in quiescent
  galaxies and $<10$ per cent in strongly starbursting galaxies.  We
  derive the fractions of all main sequence galaxies at $z<2$ that are
  experiencing a period of moderate nuclear activity, noting that it
  is strongly dependent on galaxy stellar mass (\mgal); rising from
  just a few per cent at \mgal$\sim10^{10}$~\msun\ to $\gtrsim$20 per
  cent at \mgal$\geq10^{11}$~\msun.  Our results indicate that it is
  galaxy stellar mass that is most important in dictating whether a
  galaxy hosts a moderate luminosity AGN.  We argue that the majority
  of moderate nuclear activity is fuelled by internal mechanisms
  rather than violent mergers, which suggests that high redshift disk
  instabilities could be an important AGN feeding mechanism.

\end{abstract}


\section{Introduction}
\label{Introduction}

Over the past two decades a significant bank of evidence has developed
demonstrating, to first order, the close links between the growth of
galaxies and their resident black holes (e.g., the similarity between
the black hole accretion and star formation histories of the
Universe, the tight correlation between black hole and galaxy bulge
masses; \citealt{2000ApJ...539L...9F, 2000ApJ...539L..13G,
  2004ApJ...604L..89H, 1998MNRAS.293L..49B, 1999MNRAS.310L...5F,
  2004MNRAS.353.1035M}).  Results from some theoretical models suggest
that these links arise through feedback processes between the galaxy
and its accreting black hole, regulating their respective growths
(e.g., \citealt{ 2004ApJ...600..580G, 2005Natur.433..604D,
  2006ApJS..163....1H}).  However, outside these models the physical
mechanisms driving this expected feedback are poorly understood,
primarily due to a lack of constraints from observations.  Indeed, it
is still a matter of debate whether all levels of nuclear activity
directly impact star formation (or vice-versa), or whether the scale
of their influence is a function of AGN or galaxy properties (e.g.,
AGN luminosity, galaxy mass, redshift).\footnote{Here and throughout
  we use the term ``nuclear activity'' to explicitly refer to
  accretion onto a black hole (i.e., AGN) rather than, for example,
  intense nuclear star formation.} Measuring the simultaneous levels of
nuclear and star formation activity taking place within the galaxy
population is crucial if we are to determine how or, indeed, if these
processes are causally linked.

Measuring the stellar properties of a galaxy (including star formation
activity) when faced with contamination from an AGN presents a
significant challenge.  The most commonly used diagnostics rely on
optical spectral features (e.g., the 4000~\AA\ break, the strengths of
the H$\delta_A$ absorption and $[$O~{\sc ii}$]$ emission lines), which
have been applied extensively to large scale spectral surveys to
measure the properties of large numbers of AGN host galaxies (e.g.,
\citealt{2003MNRAS.346.1055K, 2009ApJ...696..396S}).  However,
spectral diagnostics are not immune to AGN contamination and optical
diagnostics, in particular, are susceptible to the effects of
reddening.  By contrast, the far-infrared (hereafter, FIR) regime is
largely impervious to obscuration and, except in the most extreme
cases (e.g., \citealt{2011A&A...531A.137H}), provides a measure of the
star formation activity of the host galaxy that is largely
uncontaminated by the AGN (e.g., \citealt{2007ApJ...666..806N,
  2010A&A...518L..33H}).  However, it is only with the recent
availability of \Spitzer\ and, more crucially, \Herschel\ that we have
been able to measure the FIR emission of AGN hosts at high enough
redshifts to probe the epochs during which most stellar and black hole
mass in the Universe was built up (i.e., $z\gtrsim1$;
\citealt{2010MNRAS.401..995M, 2010ApJ...720..368X,
  2010A&A...518L..26S}).  These first FIR studies of AGNs in the high
redshift Universe had to contend with relatively low infrared
detection fractions (typically $\lesssim$20 per cent) and, as such,
relied heavily on stacking analyses.  By contrast, in this study we
use data from the deepest FIR survey yet undertaken by \Herschel\ to
reach a FIR detection rate of $\gtrsim$40 per cent. For the first
time, we can use individual detections to explore the trends between
AGN and star formation activity to high redshifts (i.e.,
$z\lesssim3$), only resorting to stacking analyses to confirm, rather
than identify, correlations derived from FIR observations.

The FIR output of a galaxy provides a direct measure of the rate at
which that galaxy is forming stars (i.e., the star formation rate, or
SFR; e.g., \citealt{1998ARA&A..36..189K}).  Clearly, SFRs are a
crucial element in most, if not all, studies exploring the links
between nuclear activity and galaxy growth.  However, it has recently
become apparent that the SFRs of normal star-forming galaxies are
roughly proportional to galaxy stellar mass (e.g.,
\citealt{2007ApJ...660L..43N,
  2007A&A...468...33E,2007ApJ...670..156D,2009ApJ...698L.116P,
  2009MNRAS.394....3D,2010MNRAS.401.1521M}), meaning systematic
differences in the stellar masses of difference populations of
galaxies (such as AGNs and non-AGNs) will also show as differences in
their SFRs.  {\it Specific} star-formation rates (i.e., the level of
star formation per unit stellar mass; SSFR), by contrast, take this
``mass effect'' into account, providing a measure of the relative
growth rates of galaxies.  As such, SSFRs give us a clear picture of
how rapidly a given galaxy is forming stars relative to its
star-formation history.

Specific SFRs could also help to distinguish the fuelling mechanisms
that drive both nuclear and star formation activity.  It is now
becoming increasingly evident that almost all star formation takes
place in one of two regimes: the majority ``main-sequence''
(\citealt{2007ApJ...660L..43N}) and the less common compact
``starbursts'', with the latter having significantly higher (i.e.,
factors of $\gtrsim3$) SSFRs.  Interestingly, it seems that the type
of regime is closely linked to the physical process driving the star
formation activity, with starbursts being induced by major mergers,
while internal processes (e.g., disk instabilities, turbulence) drive
main-sequence star formation (e.g., \citealt{2007A&A...468...33E,
  2008ApJ...680..246T, 2010ApJ...714L.118D, 2010MNRAS.407.2091G,
  Elbaz:2011uw}).  Since both major mergers and internal processes can
also channel gas and dust into the central regions of a galaxy (e.g.,
\citealt{1996ApJ...464..641M, 2004ARA&A..42..603K,
  2006ApJS..166....1H}), it is relevant to question how nuclear
activity fits within this regime.  Until recently, nuclear activity
was often considered to be closely linked to major merger events --
largely due to the high fractions of highly luminous, ``quasar class''
AGNs that appear to be associated with merging systems (e.g.,
\citealt{1988ApJ...325...74S, 1996ARA&A..34..749S,
  1998ApJ...492..116S, 2001ApJ...555..719C, 2010MNRAS.404..198I}).
However, a number of recent morphological studies of AGN hosts have
presented evidence to suggest that more moderate levels of nuclear
activity (i.e., sub-quasar; \lx$\lesssim10^{44}$~\ergs) are typically
associated with internally-evolving galaxies undergoing secular
evolution rather than major-mergers (e.g.,
\citealt{2005ApJ...627L..97G, 2011ApJ...727L..31S,
  2011ApJ...726...57C}).  On the other hand, measuring morphologies
and identifying evidence of mergers can be problematic, especially at
high redshifts where source faintness can introduce systematic
effects.  Specific SFRs, therefore, provide a direct, independent
means to distinguish between merger and internally induced star
formation with which to verify such claims.

In this paper, we use data from the deepest surveys yet undertaken at
FIR wavelengths (i.e., GOODS-\Herschel; hereafter, GOODS-H; P.I.:
D. Elbaz) to measure the SFRs and SSFRs of X-ray selected AGN-hosting
galaxies to $z\sim3$.  We use X-ray selection as it provides the most
accurate measure of nuclear activity, largely free from host
contamination and less affected by obscuration than, for example,
optical wavelengths.  Throughout our analyses we use the \Swift-BAT
sample of AGNs as a local comparison sample as they have similar X-ray
properties as the high redshift AGNs in the deep GOODS fields.  Here,
we largely focus on moderate luminosity AGNs (i.e.,
\lx$=10^{42}$--$10^{44}$~\ergs) which are responsible for the majority
of the integrated X-ray output of AGNs in the $z\lesssim3$ Universe
(e.g., \citealt{2003ApJ...598..886U, 2005A&A...441..417H,
  2007ApJ...654..731H, 2010MNRAS.401.2531A}).  Although, whether they
represent the majority of black hole {\it growth} depends on the
adopted bolometric correction factor.\footnote{\label{LumDepBol} The
  luminosity-dependent correction factor of \cite{2004MNRAS.351..169M}
  suggests that more luminous quasars dominate black hole growth,
  although \cite{2007MNRAS.381.1235V} argue against such a simple
  luminosity-dependent bolometric correction factor.}  Our principal
aim is to compare the SFRs and SSFRs of these moderate AGN hosts
against the general galaxy population to establish whether (a) the
levels of star formation activity around moderate luminosity AGNs
differ from the general galaxy population and (b) merger or internally
driven processes have dominated their stellar and black hole mass
build up.

This paper is structured as follows: in \S\ref{Data} we describe our
data acquisition and processing, while in \S\ref{Stacking} we describe
the stacking procedure that we use to ensure we consider the FIR
output of all X-ray AGNs in our sample, in addition to those that are
formally detected.  We describe how we convert FIR fluxes into SFRs
and SSFRs in \S\ref{Calculating} before presenting our results in
\S\ref{Results}, where we demonstrate that moderate luminosity AGNs
typically reside in main-sequence galaxies.  In \S\ref{Discussion} we
interpret these results in terms of our current understanding of
galaxy growth and consider the fraction of main-sequence galaxies
containing AGNs.  Finally, we summarise our findings in
\S\ref{Summary}.

\section{Data}
\label{Data}

Our main focus is the analysis of the FIR emission (measured by
\Herschel) of X-ray detected AGNs, as this provides a more accurate
measure of the SFRs of AGN hosts than the mid-infrared (hereafter,
MIR; e.g, \citealt{2010MNRAS.401..995M}).  However, an important
aspect of our analyses also incorporates MIR data (from \Spitzer) to
derive the average MIR to FIR SEDs of AGNs in the \Chandra\ Deep
Fields (hereafter, CDFs; see \S\ref{Results:SEDs}). We use optical and
near-infrared (hereafter, NIR) photometry to estimate host galaxy
stellar masses (hereafter, \mgal), which we combine with SFRs to
calculate SSFRs. To aid in the interpretation of the deep (i.e., high
redshift) CDF data, we make use of archival \textit{IRAS} data for the
\Swift-BAT sample of local X-ray AGNs. In this section, we describe
our X-ray AGN selection (\S\ref{Data:Xray}), the FIR, MIR, optical and
NIR data and the cross-matching between these regimes (\S\ref{Data:IR}
and \S\ref{Data:Optical}).  We end this section with a description of
the \Swift-BAT sample (\S\ref{Data:BAT}).

\begin{figure}
\begin{center}
	\includegraphics[width=8.4cm]{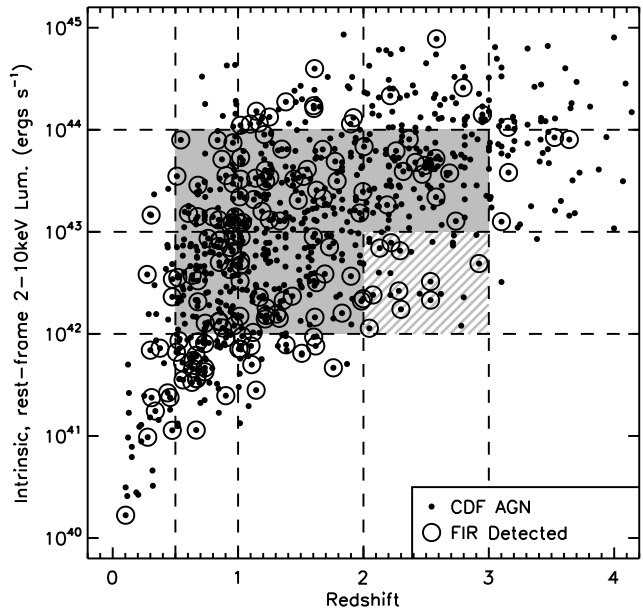}
\end{center}
\caption{Intrinsic (i.e., absorption-corrected), rest-frame 2-10~\kev\
  luminosities (i.e., \lx) of the X-ray AGNs detected by the 2Ms and
  4Ms \Chandra\ Deep Field Surveys (North and South, respectively)
  plotted as a function of redshift (see \S\ref{Data:Xray}). The AGNs
  that are detected (at $>3\sigma$) in either of the 100~\mum\ or
  160~\mum\ bands are circled (see key).  To mitigate the effect of
  biases when measuring the average infrared properties of X-ray AGNs
  in our sample we define a homogenous (in terms of \lx\ and $z$),
  restricted sample (shaded region). However, we stress caution when
  interpreting the results from the $z=2-3$,\ \
  \lx$=10^{42}$--$10^{43}$~\ergs\ bin (hatched region) due to the low
  number of X-ray detected AGNs in this redshift and \lx\ range (i.e.,
  29 X-ray AGNs).}
\label{Lxz}
\end{figure}

\subsection{X-ray data}
\label{Data:Xray}

Our initial sample selection is the 1243
\unskip
X-ray sources detected in the 2~Ms CDF-N and 4~Ms CDF-S surveys, as
reported in the \cite{2003AJ....126..539A} and
\cite{2011ApJS..195...10X} main catalogues, respectively.\footnote{We
  wish to make it clear that, throughout, we use the ``CDFs'' notation
  to refer to both \Chandra\ Deep Fields and use CDF-S to refer to the
  \Chandra\ Deep Field South only.}  The vast majority of sources in
these combined catalogues have been identified as AGNs (i.e.,
978
\unskip, or $\approx$79
\unskip per cent),
with the remainder identified as either stars (i.e.,
24
\unskip, $\approx$2
\unskip per cent) or
starbursting galaxies (i.e., 241
\unskip,
$\approx$19
\unskip per cent).  We use the classifications of
\cite{2004AJ....128.2048B} and \cite{2011ApJS..195...10X} to select
only AGNs from the CDF-N and CDF-S catalogues, respectively.  Both
classification schemes combine information derived from the optical
(spectra and/or photometry) and X-ray regimes to differentiate between
the different classes of object.  Redshifts and absorption-corrected
2-10~keV luminosities (i.e., \Lx) for each X-ray AGN in the CDF-N
catalogue were taken from the catalogue described in
\cite{2004AJ....128.2048B}.  For the CDF-S we used the values from
\cite{2011ApJS..195...10X}, converting their 0.5--8~keV to 2-10~keV
luminosities by assuming a constant conversion factor, \lx$=0.68
~L_{\rm 0.5-8keV}$, appropriate for the constant spectral index (i.e.,
$\Gamma=1.8$) used to derive their $L_{\rm 0.5-8keV}$ values.  It
should be noted that the \lx\ estimates reported by
\cite{2011ApJS..195...10X} were derived using simple band ratios to
estimate the absorbing column densities, rather than the more detailed
spectral fitting approach carried out by, for example,
\cite{2004AJ....128.2048B}.  However, \cite{2011ApJS..195...10X}
report that, when available, \lx\ values derived from spectral fitting
generally agree (i.e., to within $\sim$30 per cent) with those
estimated using band ratios (although this factor may be higher for
highly obscured sources).  Of the 978
\unskip X-ray
detected AGNs in the combined fields, 879
\unskip
(i.e., $\approx$90
\unskip per cent) have either a
spectroscopic (i.e., 480
\unskip;
$\approx$55
\unskip per cent) or photometric (i.e.,
399
\unskip; $\approx$45
\unskip per
cent) redshift measurement. The adopted X-ray luminosities have been
corrected for absorption on the basis of X-ray spectral analyses and
X-ray band ratios (\citealt{2011ApJS..195...10X}; Bauer et al. in
prep).  As a key goal of this study is to explore the infrared
properties of AGNs as a function of AGN power (i.e., \lx) and
redshift, we only consider those 879
\unskip AGNs
for which these parameters are known.  The \lx--\z\ distribution of
these AGNs is shown in \fig{Lxz}.  As for all flux limited surveys
there is a clear Malmquist bias within this sample.  Despite this, the
regions of the parameter space spanning $z=0.5-3$ and
\lx$=10^{42}$--$10^{44}$~\ergs\ are well sampled and we mainly focus
our attention on the 564
\unskip AGNs within
these ranges, although we stress caution when interpreting the results
from the single $z=2-3$, \lx$=10^{42}$--$10^{43}$~\ergs\ bin, due to
the low number of detected X-ray AGNs in this redshift and \lx\ range
(i.e., 29
\unskip X-ray AGNs).

\subsection{Mid and far-infrared data}
\label{Data:IR}

To characterise the MIR and FIR properties of our sample of X-ray
AGNs, we used the deepest data currently available at 16, 24, 100 and
160~\mum: the \Spitzer\ IRS peak-up survey (16~\mum;
\citealt{2011AJ....141....1T}; hereafter, GOODS-\IRS), the
GOODS-\Spitzer\ Legacy survey (24~\mum; P.I.:Mark Dickinson), and the
GOODS-H survey (100~\mum\ and 160~\mum; \citealt{Elbaz:2011uw}). The
16~\mum\ observations and data reduction are described in full in
\cite{2011AJ....141....1T} but, to summarise, they resulted in a pair
of mosaics with pixel scales of 0.9\arcsec\ and average 3\sig\ depths
of 40 and 65~\mujy\ in the northern and southern fields,
respectively. The 24~\mum\ observations resulted in a pair of final
mosaics with a pixel scale of 1.2\arcsec\ and a point-source 5$\sigma$
sensitivity limit of $\sim$30~\mujy\ in both fields.  The GOODS-H
observations are described in \cite{Elbaz:2011uw} and the final images
were produced following the procedure outlined in
\cite{2010A&A...518L..30B}.  In total, the GOODS-H program consists of
124.6 hours of PACS observations (\citealt{2010A&A...518L...2P}) in
the northern field, and 264 hours in the southern field (31.1 hours of
SPIRE observations were also obtained for the northern field, but we
do not use these longer wavelength data here).  The final images have
pixel scales of 1.2\arcsec\ (100~\mum) and 2.4\arcsec\ (160~\mum).  At
100~\mum\ the 5$\sigma$ sensitivities are 1.7~mJy and 1.2~mJy in the
northern and southern fields, respectively, while at 160~\mum\ they
are 4.5~mJy and 3.5-4.5~mJy, respectively.\footnote{The 100~\mum\ and
  160~\mum\ surveys cover the same region of sky.}
 
Although the various X-ray and infrared surveys are roughly centred on
the same positions on the sky (i.e., [RA, Dec]: $[12^h$:$37^m$,
$62^\circ$:$14\arcmin]$ for the northern field; $[3^h$:$32^m$,
$-27^\circ$:$48\arcmin]$ for the southern field), because of the
different fields of view and observing modes of the various
instruments they do not cover exactly the same region of the sky.
Because of this, roughly 53
\unskip,
70
\unskip and
47
\unskip per cent (i.e.,
465
\unskip,
614
\unskip,
409
\unskip) of the X-ray AGNs in our
sample with known redshifts and \lx\ are covered by the GOODS-\IRS,
GOODS-\Spitzer\ and GOODS-H surveys, respectively.  Roughly
44
\unskip per cent (i.e.,
391
\unskip) of our X-ray AGN sample is
covered by all three surveys, although we only use data from all three
when deriving the average infrared SEDs of the X-ray AGNs in our
sample (see \S\ref{Results:SEDs}); we use only FIR fluxes from the
GOODS-H survey to calculate \lir\ (see \S\ref{Calculating:LIR}).

To identify 16~\micron\ counterparts to the X-ray AGNs in the CDFs we
used the publicly available catalogue of \cite{2011AJ....141....1T}.
The fluxes reported in the 16~\mum\ catalogue were obtained by
performing PSF-fitting using the positions of 5$\sigma$ sources from
the 3.6~\micron\ GOODS catalogue as priors (see Dickinson et al. in
prep.).  The 16~\mum\ catalogue lists the fluxes of all sources
detected at $>5\sigma$ in the 16~\micron\ band and contains 1309
sources across both CDFs.  The 24~\micron\ catalogue, which is also
based on the positions of $5\sigma$, 3.6~\micron\ priors, contains
5188 sources across both CDFs at $>3\sigma$.  The 100~\micron\ and
160~\micron\ fluxes were calculated in a similar fashion by performing
PSF fitting at the positions of the $>3\sigma$ 24~\micron\ priors.  As
we are exploring general trends in the FIR properties of X-ray AGNs,
rather than focusing on individual sources, we use a relatively low
detection threshold of 3$\sigma$ at 100~\micron\ and 160~\micron\ to
maximise the number of sources in our sample.  Across both GOODS-H
fields there are 1324 and 1095 \Herschel\ sources that satisfy this
criteria at 100~\mum\ and 160~\mum, respectively.

We used positional matching to identify infrared counterparts to the
X-ray sources in each infrared band.  Because there is no significant
difference between the spatial resolutions of the 16~\mum\ and
24~\mum\ images we used the same 3.0\arcsec\ matching radius for both
wavelengths, which we noted in \cite{2010MNRAS.401..995M} provides the
best compromise between finding real matches while reducing the number
of spurious matches.  The 100~\mum\ and 160~\mum\ positions are from
24~\mum\ priors (which are, in turn, based on IRAC 3.6~\mum\ priors),
so there was no need to perform any matching for these bands.  At
16~\mum, we identified 192
\unskip matches (i.e.,
$>5$\sig\ at 16~\mum) to the X-ray detected AGNs with measured
redshifts (i.e., $\approx$41
\unskip per cent of those covered
by this survey).  At 24~\mum\ we identify
449
\unskip matches (i.e.,
$\approx$73
\unskip per cent) while at 100~\mum\ and
160~\mum\ we identify 174
\unskip (i.e.,
$\approx$43
\unskip per cent) and
140
\unskip (i.e., $\approx$34
\unskip
per cent) matches, respectively (i.e., $>3$\sig\ in each of the 24,
100 and 160~\mum\ bands). \footnote{Here, the percentages refer to the
  fraction of sources detected in the region covered by each
  individual survey, i.e., not the fraction covered by all three
  surveys.} The larger numbers of matches at 24~\mum\ compared to at
16~\mum\ is due to the increased depth of the 24~\mum\
survey. Approximately 25
\unskip per cent (i.e.,
99
\unskip) of the X-ray AGNs with measured
redshifts covered by all three surveys are detected in all four bands.
In each band the expected fraction of spurious matches is less than 2
per cent (calculated using the same procedures as outlined in
\citealt{2010MNRAS.401..995M}).

For each of the four infrared bands considered, the depth of the
observations are different in the two fields.  Furthermore, the depth
of the observations also vary as a function of position within each
field, especially in the 16~\mum\ band.  Therefore, when combining the
individual fluxes to derive mean values, we take into account these
differences by weighting as 1/error$^2$ of the measured flux.

\subsection{Optical and near-infrared data}
\label{Data:Optical}

We use optical and NIR photometry to estimate the host galaxy masses
of the X-ray AGNs in our sample (see \S\ref{Calculating:Masses}).  A
detailed description of the multi-wavelength catalog assembly and of
the photometric redshift and stellar mass estimation will be given in
a separate paper (Pannella et al. in prep.). In the following,
and for the sake of completeness, we will briefly describe the data
set and the procedures used.

In the GOODS-N field we have built a PSF-matched multi-wavelength
catalog with 10 passbands from {\it U} to 4.5~\mum. The optical, NIR
and IRAC data used here are presented in \cite{2004AJ....127..180C},
\cite{2010ApJS..187..251W} and Dickinson et al. (in prep.),
respectively. A {\it Ks}-band selected catalog has been built using
{\sc source extractor} (\citealt{1996A&AS..117..393B}) in dual image
mode. To compensate for the different resolution of the images we have
applied corrections based on the growth curve of ``bona-fide''
point-like sources. 

For the northern field photometric redshift determination closely
follows the procedure described in \cite{2009ApJ...701..787P} and
\cite{2010ApJ...714.1305S}.  When comparing to the spectroscopic
sample of \cite{2008ApJ...689..687B}, we reached a relative (i.e.,
$\Delta z = \vert z_{\rm phot}-z_{\rm spec}\vert /[1+z_{\rm spec}]$)
photo-$z$ accuracy of 4 per cent with 3 per cent of catastrophic
outliers (i.e. objects with $\Delta z > 0.2$).  For the southern field
we have used the spectroscopic or photometric redshifts reported in
\cite{2011ApJS..195...10X} throughout and used photometry from the
GOODS-MUSIC survey (\citealt{2009A&A...504..751S}) to estimate stellar
masses (to ensure consistency, we fixed the redshifts to those of
\citealt{2011ApJS..195...10X} in these mass calculations). The
GOODS-MUSIC catalog provides total magnitudes for a very similar
photometric coverage as the GOODS-N field.  We refer the reader to
\cite{2009A&A...504..751S} for a detailed description of the catalog
content.

We used positional matching to identify the counterparts to the X-ray
AGNs in the above catalogues. Rather than matching directly to the
X-ray positions themselves (which have a high positional uncertainty,
especially at large off-axis angles), we used the positions of their
optical counterparts reported in \cite{2003AJ....126..632B} and
\cite{2011ApJS..195...10X} for the northern and southern fields, respectively.
We use a matching radius of 0.5\arcsec in both fields and match
$\approx$95
\unskip per cent (i.e.,
388
\unskip) of the
409
\unskip X-ray AGNs with redshift
measurements and GOODS-H coverage.

We found that for 27
\unskip (i.e.,
$\approx$ 7
\unskip per cent of the total sample covered by
GOODS-H) AGNs in the northern field the redshift from the
\cite{2003AJ....126..632B} differed from our own new estimate by a
significant amount (i.e., $\Delta z>0.2$).  We reject these sources
entirely from our analyses as we believe that the
\cite{2003AJ....126..632B} photometric redshift is now out-of-date
(i.e., has been superseded due to improved photometric/spectroscopic
data) meaning the value of \lx\ is likely incorrect.  This ensures
that, across both fields, we have used fully consistent redshifts
throughout to derive \lx, \lir\ and \mgal.

\subsection{The {\it Swift}--BAT sample of $z\sim0$ AGNs}
\label{Data:BAT}

As we described in \cite{2010MNRAS.401..995M}, the \Swift-BAT sample
of AGNs (\citealt{2008ApJ...681..113T}) is currently one of the best
local comparison samples for studies of high redshift, X-ray AGNs
detected in the CDFs (see Figs. 7 and 10 of
\citealt{2009ApJ...690.1322W} where they plot the \lx\ and \nh\ of the
\Swift-BAT AGN sample).  Of the 104 \Swift-BAT AGNs for which \lx\ and
\nh\ have been measured (\citealt{2009ApJ...690.1322W}), 41 have
infrared flux density measurements from all four \IRAS\ bands (these
41 objects are hereafter referred to as the BAT/\IRAS\ sample),
allowing direct comparison with the infrared properties of the AGNs in
the CDFs.  All of these AGNs have $z<0.1$ and are thus ``local'' in
comparison to the $z=0.5$--3 CDF AGNs which form our high redshift
sample.  As none of the Blazars in the \cite{2008ApJ...681..113T}
  have \lx\ measurement reported in \cite{2009ApJ...690.1322W} they
  are automatically excluded from our sample.

\section{Mid and far-infrared stacking procedure}
\label{Stacking}

By using the deepest FIR observations currently available, we detect
roughly $\approx$46
\unskip per cent of X-ray sources
at $>3\sigma$ in either the 100~\mum\ or 160~\mum\ bands.  This
represents a vast improvement over previous FIR surveys (e.g.,
$\approx11$ and $\approx20$ per cent for \citealt{2010MNRAS.401..995M}
and \citealt{2010A&A...518L..26S}), respectively. However, we resort
to stacking techniques (see \S\ref{Results:LIR:Undetected}) to confirm
that the trends observed in the FIR-detected fraction of our sample
remain when {\it all} X-ray AGNs in our sample are considered (i.e.,
even those undetected at FIR wavelengths).

As our intention is to explore how the infrared properties (e.g.,
infrared luminosities, SED shape) change as a function of redshift and
nuclear activity, we stack subsamples of our X-ray AGNs separated into
six different \lx\ and \z\ bins: all combinations of $z=$0.5--1, 1--2,
2--3 and \lx$=10^{42}$--$10^{43}$~\ergs and
$10^{43}$--$10^{44}$~\ergs; the number of AGNs in each stack are given
in column 4 of table \ref{TableAvg}.  These bins cover the region of
the \lx--\z\ parameter space that benefits from high levels of
completeness (see \S\ref{Data:Xray} and \fig{Lxz}).  The nature of
small, yet deep fields such as the CDFs implies that, while we can
probe moderate luminosity (i.e., \lx$=10^{42}$--$10^{44}$~\ergs) AGNs
to high redshifts, there are only a small number of the rarer, quasar
luminosity AGNs in our sample, especially at lower redshifts (i.e.,
$z\lesssim2$).  Wide, but less deep, infrared surveys such as PEP
(\citealt{2011A&A...532A..90L}) and HERMES (P.I.: S. Oliver) are
needed to fully explore the average infrared properties of both higher
luminosity (i.e., \lx$>10^{44}$~\ergs) or, alternatively, lower
redshift (i.e., $z\lesssim0.5$) AGNs.

We only stack on residual images (i.e., images from which the sources
down to 3\sig\ have been removed), stacking only those X-ray sources
that are undetected in the various infrared bands (i.e., $<5\sigma$
for 16~\mum\ and $<3\sigma$ for 24~\mum, 100~\mum\ and 160~\mum).  We
take this approach, rather than stacking all X-ray positions on the
original images, to reduce the effect of flux boosting from nearby
bright sources.  However, as the residuals from very bright sources
can be strong even after source subtraction, we do not stack at
positions that lie within 10\arcsec\ of $>5\sigma$ sources
(equivalently, $>$65--110~\mujy, $>$1.2--1.7~mJy and $>$4--4.5~mJy at
16~\mum, 100~\mum\ and 160~\mum, respectively).  The stacked images
are produced by summing cut-out images centred on the positions of the
infrared-undetected X-ray sources in both fields.  Prior to summation,
each pixel in the individual cutouts is weighted by the inverse of its
RMS noise, determined from RMS maps of the fields.  The weighted mean
stack is then calculated by dividing the sum of these cutouts by the
sum of the weights on a pixel-by-pixel basis, i.e.,

\begin{equation}
\label{stackeqn}
\overline{S}(x,y) = \frac{\sum_{i=1}^{n} S_i(x,y)w_i(x,y)}{\sum_{i=1}^{n} w_i(x,y)}
\end{equation}

\noindent
where $\overline{S}_(x,y)$ is the final stacked image, $S_i(x,y)$ and
$w_i(x,y)$($=1/{\rm RMS^2}[x,y]$) represent the individual cutout
images and weight map, respectively, and $n$ is the total number of
cutouts to be combined in the stack.  Once the weighted mean stacks
have been produced for the sources in each of our bins, we use
aperture photometry to measure the flux within a circular aperture
placed at the centre of each stacked image. The radius of this
aperture depends on the PSF of the observation and thus, ultimately,
on the wavelength and instrument: 6\arcsec\ for both the 16~\mum\ and
24~\mum\ wavebands and 9\arcsec\ and 15\arcsec\ at 100~\mum\ and
160~\mum, respectively.  Through simulated stacks, we find that the
100~\mum\ and 160~\mum\ fluxes extracted through aperture photometry
are 70 per cent of those derived from PSF fitting (i.e., the technique
used to extract individually detected sources).  To correct for this
discrepancy, we apply an aperture correction to our stacked fluxes,
such that the reported flux is $0.7^{-1}$ times the flux derived from
the stacks.  We also include a 15 per cent upwards correction factor
to account for the effects of source masking that was performed when
removing the low-frequency noise from the raw \Herschel\ images.  We
then use a weighted mean (here, weighting by number) to combine this
flux derived from stacking with the mean flux from the individually
detected sources.  We follow the same procedure (aside from minor
modifications to account for the different image resolutions) to stack
in each of the four infrared bands.

The errors on the fluxes derived from these stacks of the undetected
sources are calculated by using the same procedure to stack the same
number of random positions in the residual frames. We perform this
``random stacking'' 10,000 times, with the error calculated by taking
the standard deviation of the fluxes from these trials.

We use a different bootstrapping approach to calculate the errors on
the total mean 100~\mum\ and 160~\mum\ fluxes (i.e., those derived by
combining the detected and undetected mean fluxes). This ``error on
the mean'' is different from both the error on the stacked fluxes and
the standard deviation of the samples.  Taking each bin in turn, we
randomly select (with replacement) $^1/_3$ of the sample within that
bin, taking no account of whether they are detected or not, and use
the procedure outlined above to calculate the weighted mean flux of
this $^1/_3$ sample. This is performed 10,000 times for each bin and
the error on the mean is calculated by taking $\sqrt{^1/_3}$ times the
standard deviation of the results of these trials.

\section{Calculating total infrared luminosities, star formation rates,
  host galaxy masses and specific star formation rates}
\label{Calculating}

One of the main aims of this study is to relate nuclear activity to
star formation at $z\lesssim3$.  A key measurement is therefore the
SFRs and SSFRs of the AGN hosts in our sample.  In this section we
describe how we measure \lir\ for the AGN hosts in our sample and
convert these into star formation rates.  We also describe how we use
deep optical and NIR observations to measure the stellar masses of
these galaxies, enabling us to calculate their SSFRs.

\begin{figure}
\begin{center}
	\includegraphics[width=8.4cm]{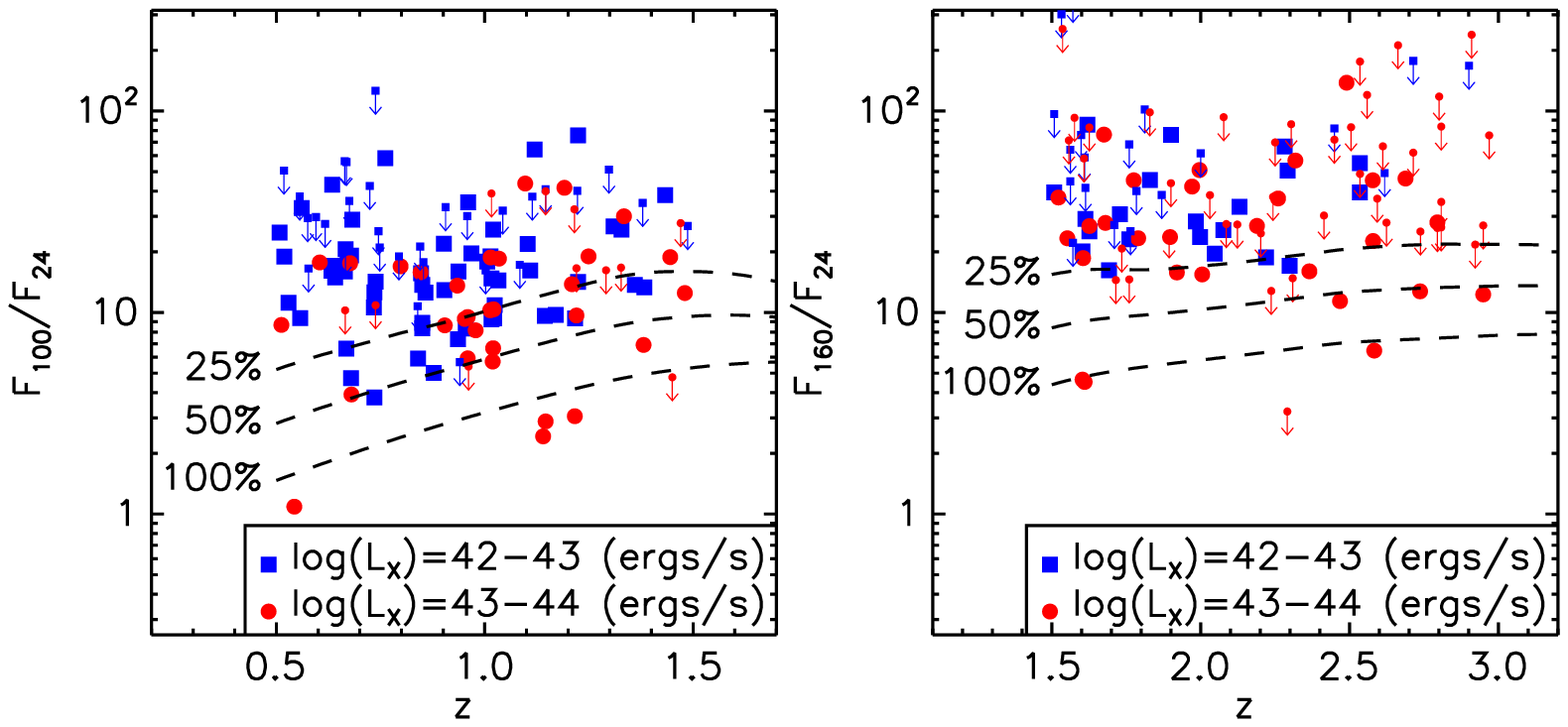}
\end{center}
\caption{Observed $[$100~\mum$/$24~\mum$]$ ({\it left}) and
  $[$160~\mum$/$24~\mum$]$ ({\it right}) flux ratios vs. source
  redshift for the X-ray AGNs in our sample.  Blue squares and red
  circles represent \lx$<10^{43}$~\ergs\ and $>10^{43}$~\ergs\ AGNs,
  respectively.  Downward pointing arrows indicate 3\sig\ upper limits
  for those X-ray AGNs that are undetected at FIR wavelengths. We use
  the observed 100~\mum\ and 160~\mum\ fluxes to measure \lir\ for
  AGNs at $z=0.5-1.5$ and $z=1.5-3$, respectively (see text,
  \S\ref{Calculating:LIR}).  Included in these plots are tracks
  indicating the expected flux ratios if 25, 50 and 100 per cent of
  the flux in the 100~\mum\ and 160~\mum\ bands arises from the AGN
  (estimated using the empirical SED templates of \protect
  \citealt{2011MNRAS.414.1082M}).  We exclude from further
  consideration any systems in which these ratios indicate that more
  than 50 per cent of the 100~\mum\ or 160~\mum\ flux (depending on
  redshift) is dominated by the AGN (16 in total, see text;
  \S\ref{Calculating:LIR}).}
\label{FIRz}
\end{figure}

\subsection{Total infrared luminosities and star formation rates}

\label{Calculating:LIR}
We use the SED library of \cite{2001ApJ...556..562C} to calculate the
integrated 8-1000~\mum\ infrared luminosities (hereafter, \lir) of the
AGN hosts in our sample.  These SEDs, when combined with the
monochromatic luminosities measured by the 100~\mum\ and 160~\mum\
\Herschel-PACS bands, have been shown to provide reliable estimates of
\lir\ at the redshifts considered here (e.g.,
\citealt{2010A&A...518L..29E}).  The shapes of the
\cite{2001ApJ...556..562C} SEDs are purely a function of infrared
luminosity, meaning that each SED has a single characteristic \lir\
(and vice-versa).  We identify the SED that most closely matches the
rest-frame monochromatic luminosity (i.e., $\nu L_{\nu}$) derived from
either the observed 100~\mum\ or 160~\mum\ flux then integrate this
SED between 8-1000~\mum. The choice of FIR waveband depends on the
redshift of the source; for $z=$0.5--1.5 sources, we use the 100~\mum\
band, while for $z=$1.5--3.0 sources, we use the 160~\mum\ band.  This
ensures that we always sample roughly the same rest-frame wavelengths
(i.e., 40--67~\mum), thus minimising the systematic uncertainties
introduced by $K$-corrections.  Roughly
$\approx$42
\unskip per cent (i.e.,
144
\unskip) of the $z=0.5-3$ X-ray AGNs in
our sample that are covered by the GOODS-H footprint are detected in
their respective FIR bands; hereafter, we collectively refer to these
as being ``FIR detected''.

\begin{figure*}
\begin{center}
	\includegraphics[width=14.0cm]{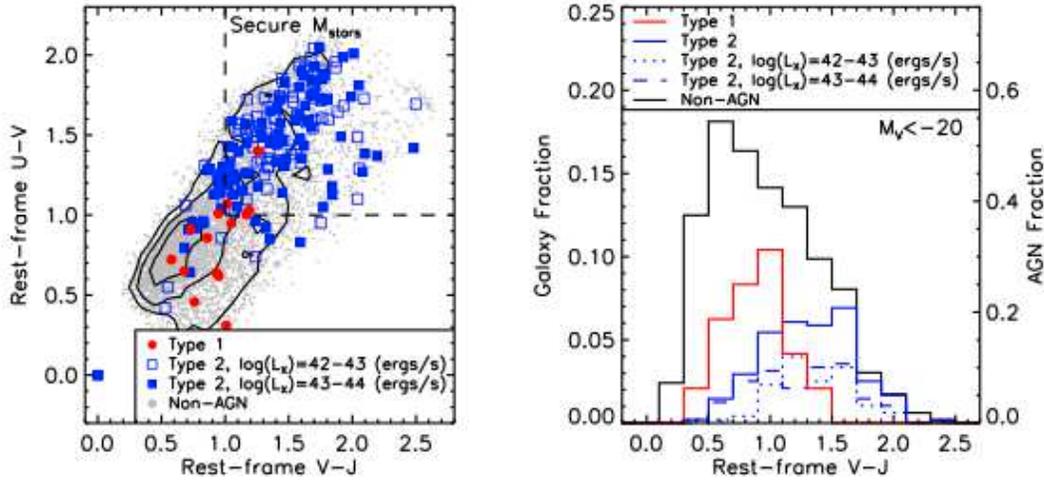}
\end{center}
\caption{{\it Left}: Rest-frame $U-V$ vs. $V-J$ colours for the AGNs
  in our sample, over-plotted onto the colour-colour distribution of
  non-AGN galaxies in the same fields (grey points).  We use contours
  to show the density of the non-AGNs in this plot.  In general, type
  2 AGNs tend to be {\it redder} than the galaxy population, which we
  argue indicates that their rest-frame optical to NIR SEDs (which we
  use to estimate host galaxy stellar masses, \mgal) are dominated by
  the host galaxy, rather than the AGN (see
  \S\ref{Calculating:Masses}).  As a precaution, we only use host
  galaxy stellar masses in cases where $U-V>1.0$ and $V-J>1.0$
  (indicated by the dashed lines). {\it Right}: Histogram showing the
  distribution of $V-J$ colours for both Type 1 and Type 2 AGNs in our
  sample and the underlying galaxy population.  For clarity we have
  expanded vertical scale of the galaxy fraction by a factor of 3.}
\label{V_min_J}
\end{figure*}

\begin{figure*}
\begin{center}
	\includegraphics[width=14.0cm]{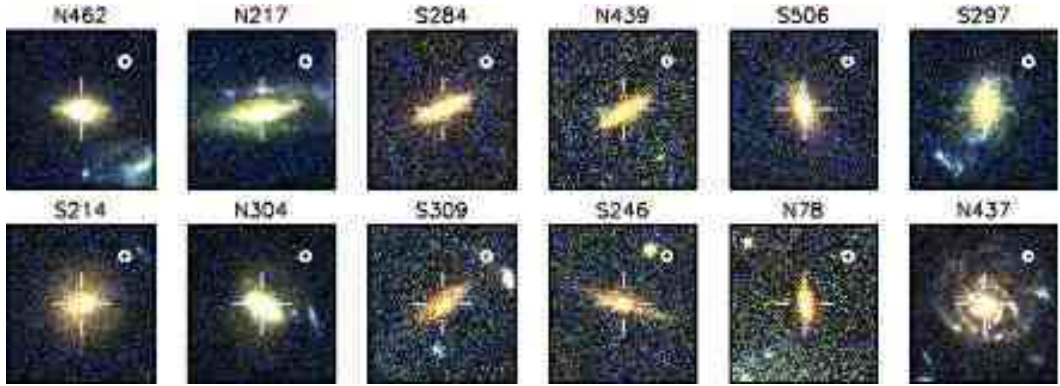}
\end{center}
\caption{Example {\it Hubble Space Telescope} ACS three-colour (i.e.,
  {\it B, V, I}) images of the galaxies identified as having
  ``secure'' stellar mass measurements (see
  \S\ref{Calculating:Masses}).  We primarily use colour-colour
  selection to identify those galaxies whose integrated light is
  likely dominated by the host-galaxy (and therefore gives a reliable
  measure of the stellar mass of the host galaxy).  As confirmation of
  this, we note that $>90$ per cent of $z<1$ galaxies selected in this
  way also show clear evidence of having extended morphologies and/or
  are not dominated by a central, point-like source (i.e., AGN).  At
  higher redshifts it becomes increasingly difficult to assess galaxy
  morphologies using optical imaging, so we rely solely on rest-frame
  colour-colour selection at $z>1$.  The diameter of the small circle
  in the top-right of each image corresponds to the spatial resolution
  of the instrument in the {\it I}-band.  The labels are a combination
  of the field (N=CDFN, S=CDFS) and the index or the X-ray source from
  either \protect \citeauthor{2003AJ....126..539A} (\protect
  \citeyear{2003AJ....126..539A}; CDFN) or \protect
  \citeauthor{2011ApJS..195...10X} (\citeyear{2011ApJS..195...10X};
  CDFS).  The crosshairs indicate the position of the optical
  counterpart to the X-ray source.}
\label{HST}
\end{figure*}

As the \cite{2001ApJ...556..562C} catalogue only includes templates
for non-AGN-hosting galaxies, we must ensure that these rest-frame
wavelengths are not dominated by the AGN, otherwise the \lir\ of the
host galaxies will be overestimated. For this, we calculate the
observed-frame $[$100~\mum$/$24~\mum$]$ and $[$160~\mum$/$24~\mum$]$
flux ratios for the $z=$0.5--1.5 and 1.5--3.0 AGNs, respectively,
detected in these bands. These ratios are presented in the left and
right panels of \fig{FIRz}. Included in these plots are tracks
indicating the expected ratios if 25, 50 and 100 per cent of the flux
in the 100~\mum\ and 160~\mum\ bands arises from the AGN. These tracks
are calculated by combining the empirical average-AGN and ``SB5'' SED
templates from \cite{2011MNRAS.414.1082M} and are appropriate for the
\lx\ and \lir\ ranges considered here. We only calculate \lir\ when at
least $50$ per cent of the 100~\mum\ or 160~\mum\ flux (depending on
redshift) is attributable to the host galaxy. The tracks shown in
\fig{FIRz} are largely unaffected by emission or absorption features
in the host galaxy templates because the 24~\mum\ band is always
dominated (i.e., $>70$ per cent) by the AGN when $<75$ per cent of the
observed 100~\mum\ and 160~\mum\ flux arises from the host
galaxy. That said, the levels of extinction applied to the AGN
component can alter these tracks. When the average AGN SED from
\cite{2011MNRAS.414.1082M} is used (which already includes a modest
amount of extinction) we remove a total of
16
\unskip FIR-detected X-ray AGNs from our high
redshift (i.e., CDF) sample (i.e., $\approx$6
\unskip per
cent of the sample). If, instead, we were to use a heavily
extinguished AGN component (such as that of ESO103--035 or NGC~5506;
see \citealt{2011MNRAS.414.1082M}) the number of rejected sourced
would increase by 18 (i.e., by a factor of $\approx2$), but this would
not affect our overall conclusions.  We identify 16 of the 41
BAT/\IRAS\ AGNs (i.e., $\approx$40 per cent) with 60~\mum\ flux likely
dominated by the AGN, rather than the host galaxy.  Since such a high
fraction of the BAT/\IRAS\ comparison sample is found to be AGN
dominated, removing them entirely from our analyses would give a false
impression of the average \lir\ and star-formation activity of local
AGNs (especially at \lx$>10^{43}$~\ergs, where two-thirds of
BAT/\IRAS\ AGNs are ``AGN dominated'').  Instead, we assume that their
rest-frame 60~\mum\ fluxes set a hard upper limit on the infrared
luminosities of their host galaxies.  Star formation rates are
calculated from the host galaxy infrared luminosities using the
prescription outlined in \cite{1998ARA&A..36..189K} assuming a
\cite{1955ApJ...121..161S} IMF, i.e.,
\begin{equation} \frac{{\rm SFR}}{{\rm M_\odot~yr^{-1}}} =
  1.7\times10^{-10}\left(\frac{L_{\rm IR}}{{\rm L_\odot}}\right)
\end{equation}
\noindent
and assuming \lsun$=3.8\times10^{33}$~\ergs.

From the rest-frame ultraviolet luminosities we estimate that
including the contribution from unobscured star formation (as in e.g.,
\citealt{2007ApJ...670..156D}) would have a negligible effect on our
results (i.e., at the level of $1.0^{+1.8}_{-0.7}$ per cent) and, for
simplicity, is neglected from the rest of our analyses.

We use the same procedure to calculate the mean \lir\ and SFRs of the
AGN hosts by treating the mean 100~\mum\ or 160~\mum\ flux (derived
from stacking analyses) as an individual source at the mean redshift
of the subsample.  Errors on these average \lir\ and SFRs are
calculated by passing the average FIR flux $\pm~1\sigma$ (see
\S\ref{Stacking}) through this process.

\subsection{Host galaxy stellar masses and specific star formation
  rates}

\label{Calculating:Masses}
Stellar masses were estimated for both fields using the SED fitting
code described in detail in \cite{2004ApJ...616L.103D} and
\cite{2009ApJ...707.1595D}.  We parametrize the possible star
formation histories (SFHs) by a two-component model, consisting of a
main, smooth component described by an exponentially declining star
formation rate $\psi(t) \propto \exp(-t/\tau)$, linearly combined with
a secondary burst of star formation. The main component timescale
$\tau$~varies ~in $\in [0.1,20]$~Gyr, and its metallicity is fixed to
solar. The age of the main component, $t$, is allowed to vary between
0.01~Gyr and the age of the Universe at the object's redshift.  The
secondary burst of star formation, which cannot contain more than 10\%
of the galaxy's total stellar mass, is modelled as a 100~Myr old
constant star formation rate episode of solar metallicity.  We adopt a
\cite{1955ApJ...121..161S} initial mass function for both components,
with lower and upper mass cutoffs of 0.1 and 100~\msun.

Adopting the \cite{2000ApJ...533..682C} extinction law, both the main
component and the burst are allowed to exhibit a variable amount of
attenuation by dust with $A_V^{1,2}$~$\in [0,1.5]$~and~$[0, 2]$ for
the main component and the burst, respectively.  For the GOODS-N
field, we have corrected the ``aperture magnitude'' stellar masses to
total stellar masses by means of the ratio of total (FLUX\_AUTO) and
aperture fluxes in the $K$-band detection image.  Of the
346
\unskip host-dominated X-ray AGNs with
GOODS-H coverage and consistent redshifts (see \S\ref{Data:Optical}),
we have been able to calculate stellar masses for
327
\unskip (i.e.,
$\approx$95
\unskip per cent).

Of course, our mass estimates are based on the assumption that the
light in these optical and NIR bands is dominated by the host galaxy,
rather than the AGN.  While a number of previous studies have found
that this is indeed the case for \lx$<10^{44}$~\ergs\ AGNs such as
those considered here (e.g. \citealt{2008ApJ...675.1025S,
  2008ApJ...681..931B, 2010ApJ...720..368X}), we wanted to perform our
own tests to confirm that this is true for our sample.

\begin{figure*}
\begin{center}
	\includegraphics[width=14.0cm]{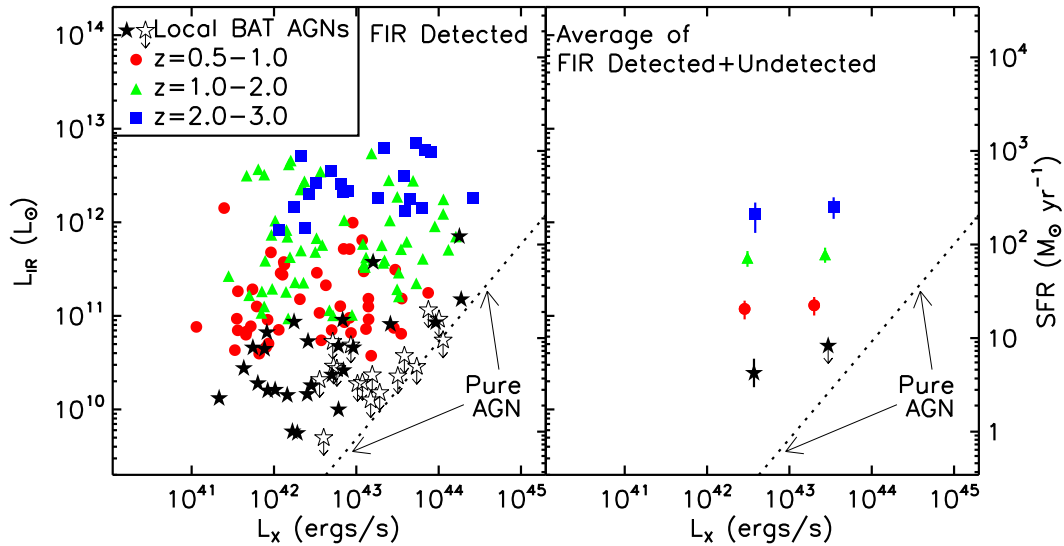}
\end{center}
\caption{{\it Left}: Total infrared luminosities of X-ray AGNs
  vs. intrinsic (i.e., absorption-corrected) rest-frame 2-10~\kev\
  luminosities (i.e., \lx) for the far-infrared detected X-ray AGNs in
  our sample.  We have separated the sample in terms of redshift (see
  key).  Open stars are used to indicate those AGNs in the local
  (i.e., $z<0.1$) BAT/\IRAS\ comparison sample whose 60~\mum\ flux is
  likely dominated by emission from the AGN, rather than the host
  galaxy.  In such cases the \lir\ shown here can be considered a hard
  upper limit on the infrared luminosity of the host galaxy (see
  \S\ref{Calculating:LIR}).  In a given redshift bin there is no
  significant correlation between X-ray and infrared luminosities, but
  we see a strong increase in infrared luminosity with increasing
  redshift.  {\it Right}: Same as {\it left}, but here we show the
  average infrared luminosities derived from stacking analyses (see
  text, \S\ref{Stacking}) vs. mean-average \lx\ for each of our \lx\
  and redshift bins. Since two-thirds of
  \lx$=10^{43}$--$10^{44}$~\ergs\ AGNs from the BAT/\IRAS\ comparison
  sample only have upper limits on their host galaxy \lir, we estimate
  an upper limit for the average value.  Again, we see a clear
  increase in \lir\ with redshift, but no significant change over the
  \lx$=10^{42}-10^{44}$~\ergs\ range considered here.  In both panels
  we indicate the \lir\ expected from a ``pure'' AGN (i.e., without
  any host-galaxy contamination; dotted line), calculated using Eq. 4
  of \protect \cite{2011MNRAS.414.1082M}.}
\label{LIRLx}
\end{figure*}

To test whether the optical/NIR fluxes of our AGNs are host-dominated
we compare their rest-frame colours against those of the general
galaxy population.  The central engines of AGNs have blue optical--NIR
colours as their broad-band UV to NIR SEDs rise monotonically to
shorter wavelength - a result of this light being emitted directly
from the accretion disk itself. Therefore, the presence of a
dominating AGN will tend to result in the galaxy as a whole (i.e.,
stellar + AGN components) having a bluer colour than if the AGN were
not present.  In the left-hand panel of \fig{V_min_J} we plot the
rest-frame $U-V$ and $V-J$ colours of the AGNs in our sample,
separated into Type 1 (i.e., optically unobscured; broad-line) and
Type 2 (i.e., optically obscured; narrow-line)
AGNs.\footnote{Rest-frame colours are calculated in the
  \cite{1978A&A....70..555B} $U$ and $V$ filters and in the $J$ band
  transmission curve of the WIRCam instrument.}  Also included in this
plot are the rest-frame colours of the underlying galaxy population in
the GOODS fields.  Rather than having particularly blue colours, Type
2 AGNs tend to lie among the redder galaxies in this colour-colour
plot, indicating that their active nuclei are obscured and that their
optical-NIR SEDs are likely host-dominated.  Furthermore, there is no
evidence to suggest that more X-ray luminous Type 2 AGNs have
systematically different $U-V$ and $V-J$ colours.  This is more
clearly shown in the right-hand panel of \fig{V_min_J}, where we plot
histograms of the $V-J$ colours of Type 1 and Type 2 (separated
according to \lx) AGNs and the underlying galaxy population.  Type 1
galaxies, on the other hand, tend to have much bluer colours, as may
be expected if the unobscured light from the accretion disk is
dominating the output of the total galaxy.  As a precaution, we only
use galaxy stellar masses for those AGNs in our sample with rest-frame
$U-V>1.0$ and $V-J>1.0$.  This leaves 255
\unskip AGNs
in the GOODS-H fields with ``secure'' masses, of which
115
\unskip are FIR detected.  However, we note
that our conclusions regarding the host galaxy stellar masses and
SSFRs do not change if we, instead, include all AGNs in our analyses.
We note that $>94$ per cent of the AGNs in our sample that satisfy
these criteria also satisfy $\nu_k L_k$/$\nu_X$\lx$>2$, which
\cite{2011MNRAS.410.1174B} suggest is a reliable indicator of a
host-dominated optical-NIR SED.  As a final test, we exploit {\it
  Hubble Space Telescope B, V, I, z} (i.e., ACS) images to check the
optical morphologies of the selected galaxies to ensure that they are
not dominated by a bright, central point source (i.e., AGN; see
\citealt{2011arXiv1107.2147A} for a detailed discussion on this
point).\footnote{Images were retrieved via the ACS GOODS Cutout
  Service v2.0: http://archive.stsci.edu/eidol\_v2.php} We find that
$>90$ per cent of $z<1$ galaxies in our sample have clear, extended
morphologies (see \fig{HST}), indicating that their integrated light
is dominated by the host galaxy, rather than the AGN.  At higher
redshifts, it becomes increasingly difficult to assess optical
morphologies, so we assume that our rest-frame colour cuts provide the
same success rate at all redshifts.  However, we also note that recent
studies exploiting new NIR data from WFCAM3 on-board {\it Hubble}
(i.e., the CANDELS survey; P.I.s: H. Ferguson, S. Faber) find that
AGN-hosts at these high redshifts have quantitatively similar
morphologies to their non-AGN counterparts, again suggesting that
their integrated light is dominated by the host galaxy (D. Kocevski,
private communication).

We calculate mean stellar masses for each of our \lx\ and $z$ bins and
use bootstrapping to calculate the error on these means in a manner
consistent with that use to determine the errors on our mean 100~\mum\
and 160~\mum\ fluxes (and hence, \lir\ and SFRs; see
\S\ref{Stacking}).  Taking each bin in turn, we randomly select (with
replacement) $^1/_3$ of the sample within that bin, and calculate the
mean stellar mass of this $^1/_3$ sample. This is performed 10,000
times for each bin and the error on the mean stellar mass is
calculated by taking $\sqrt{^1/_3}$ times the standard deviation of
the results of these trials. Errors on the SSFRs are derived by
combining the errors on the SFRs and stellar masses in quadrature.
Mean stellar masses and their associated errors for the AGNs in each
of our \lx\ and redshift bins are given in table \ref{TableAvg}.

\section{Results}
\label{Results}

In this section we present our results and explore the links between
nuclear activity, measured from \lx, and the levels of star formation
activity, measured from \lir, taking place in their host galaxies.  We
also pay particular attention to how star formation activity in AGN
host galaxies changes with redshift.  First, we consider the
$\approx40$ per cent of X-ray AGNs for which we are able to calculate
host galaxy \lir\ values and SFRs, noting that we have already
excluded any galaxies where the AGN likely dominates the observed FIR
flux (see \S\ref{Calculating:LIR}).  Then, we use the results from our
stacking analyses to confirm that trends seen in this FIR-detected
fraction of our sample extend to the general X-ray AGN population.

\subsection{Infrared luminosities and star formation rates of
  AGN-hosting galaxies}
\label{Results:LIR}

\subsubsection{X-ray AGNs detected at 100~$\mu m$ and 160~$\mu m$}
\label{Results:LIR:Detected}

In the left-hand panel of \fig{LIRLx} we plot \lir\ versus \lx\ for
all the FIR-detected AGNs in both the high redshift CDFs and local
BAT/\IRAS\ samples. Over the entire sample, which spans over three
orders of magnitude in both \lir\ and \lx, we find no statistically
significant correlation between \lir\ and \lx\ (Spearman's rank
correlation coefficient: 0.29
\unskip).  Furthermore, this
\lir--\lx\ correlation typically gets even weaker when we consider the
CDF AGNs in discrete redshift bins (i.e., $z=$0.5--1, 1--2, 2--3;
Spearman's rank: 0.16
\unskip, 0.09
\unskip and
0.30
\unskip, respectively).  We note that there is a slightly
stronger \lir--\lx\ correlation within the BAT/\IRAS\ sample
(Spearman's rank: 0.36
\unskip) due to the rise in \lir\ of
the BAT sample at \lx$\gtrsim10^{43}$~\ergs.  However, this is likely
due to the large fraction of the AGNs in this low redshift sample
BAT/\IRAS\ being AGN dominated at 60~\mum\ (see
\S\ref{Calculating:LIR}).  Indeed, the minimum \lir\ produced by a
``pure'' AGN increases almost linearly with \lx\
(\citealt{2011MNRAS.414.1082M}; indicated by the dotted line in
\fig{LIRLx}).  By splitting the sample into redshift bins, we find
that, generally, the \lir\ of AGNs increase strongly with redshift.
Indeed, the AGNs in our highest redshift bin (i.e., $z=$2--3)
typically have \lir\ and, correspondingly, SFRs around 1.5 orders of
magnitude higher than the $z<0.1$ BAT/\IRAS\ AGNs {\it with the same
  X-ray luminosities} (see \S\ref{Results:LIR:Undetected} for a more
quantitative breakdown of the increase in the average \lir\ with
redshift).  The weak correlation in the unbinned sample arises from a
combination of this strong \lir--$z$ correlation and an increase in
the proportions of more X-ray luminous AGNs in our sample at higher
redshifts; a result of both selection effects and a real, observed
increase in the average \lx\ with increasing redshift (e.g.,
\citealt{2005AJ....129..578B, 2005A&A...441..417H}).

\begin{figure*}
\begin{center}
	\includegraphics[width=14.0cm]{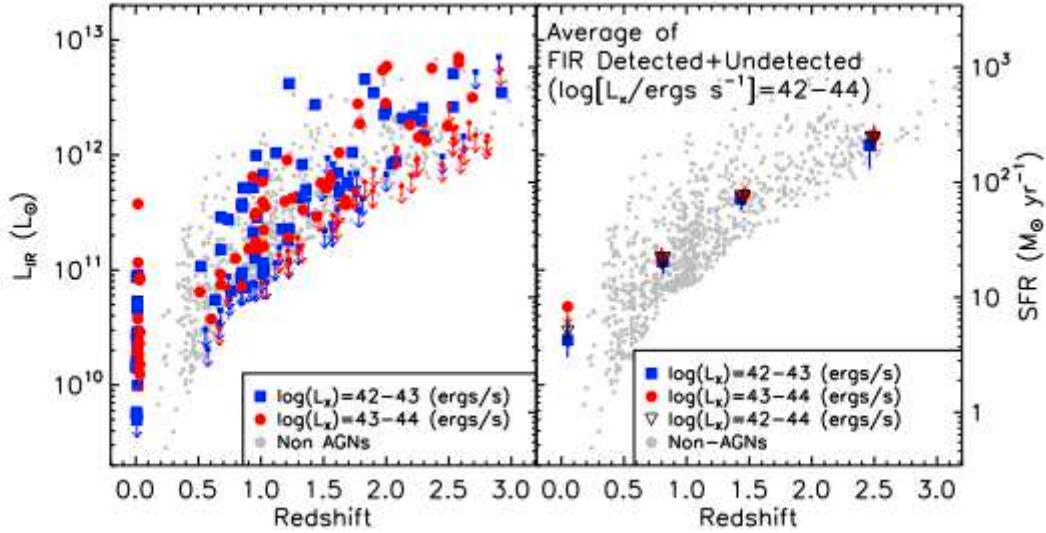}
\end{center}
\caption{{\it Left:} Rest-frame infrared luminosities vs. redshift for
  our sample of X-ray AGNs, split in terms of intrinsic, rest-frame
  2--10~keV luminosities (i.e., \lx; see key).  Downward pointing are
  used to indicate the 3\sig\ upper on \lir\ in cases where there is
  no formal FIR-detection. The points around $z\approx0$ represent the
  BAT/\IRAS\ sample of local AGNs, which cover roughly the same range
  of X-ray luminosities as the high redshift AGNs in the \Chandra\
  deep fields. While there is a clear Malmquist bias, the increase in
  the upper envelope of the distribution with redshift is too large to
  be explained in terms of the increase in the surveyed comoving
  volume with increasing redshift (see \S\ref{Results:LIR:Detected}).
  The distribution of AGNs in this plane is similar to those of
  FIR-detected non-AGN hosts (shown as grey points in both panels)
  {\it Right:} The same as the {\it left} panel, but now showing the
  average \lir\ (derived from our stacking analyses; see
  \S\ref{Stacking}) and \lx\ values.  Again, the increase in \lir, and
  therefore host galaxy star formation rates (shown by the right hand
  axes), is clear (see \S\ref{Results:LIR:Undetected}).}
\label{LIRz}
\end{figure*}

The increase in \lir\ with increasing redshift is more clearly visible
in \fig{LIRz}, where we plot \lir\ as a function of redshift.  Here,
we have included upper limits on \lir, calculated in the same way as
for the detected sources assuming 3$\sigma$ upper limits on the FIR
fluxes.  Qualitatively, the distribution of AGNs in the \lir--$z$
plane is the same as that of the FIR-detected non-AGNs in the same
field.  As is expected when $\approx$60 per cent of a sample is not
detected, this plot shows a clear Malmquist bias.  However, it is
clear that the upper envelope of the \lir-$z$ relation increases
strongly with redshift; by about 1.5 orders of magnitude from $z=0.5$
to 3.0, i.e., \lir$\sim$2\e{11} to $10^{13}$~\lsun.  This trend
continues to the low redshifts (i.e., $z < 0.1$) spanned by the BAT
sample and is too steep to be explained in terms of the increase in
the comoving volume of the survey expanding the range of \lir\ at
higher redshifts; the probed comoving volume between $z=2-3$ is only a
factor of 4 times larger than that between $z=0.5-1$.

We have demonstrated that the FIR-detected AGNs in both the high
redshift CDF and low redshift BAT samples show no strong increase in
\lir\ with \lx, yet are consistent with a strong increase in \lir\
with increasing redshift.  We will look more closely into the causes
and consequences of these results in \S\ref{Results:Masses} and
\S\ref{Discussion} but, first, we will confirm whether these results
still hold when we consider {\it all} X-ray AGNs in the GOODS fields,
i.e., not just those detected at FIR wavelengths.  For this, we resort
to stacking analyses.

\subsubsection{Results from stacking analyses}
\label{Results:LIR:Undetected}

By considering only the detected sources in the previous subsection,
we sampled only the most intrinsically infrared luminous galaxies at
each redshift.  In this subsection we consider the contribution of
X-ray AGNs that are not formally detected in each of the wavebands by
relying on stacking analyses (see \S\ref{Stacking}) to push the
average detection threshold beyond what is achievable for individual
sources.

In the right-hand panel of \fig{LIRLx} we plot the average \lir\
(derived from stacking analyses) vs. average \lx\ for CDF AGNs in each
of our two \lx\ bins and three redshift bins (i.e., $L_{\rm
  X}=10^{42}$--$10^{43}$~\ergs\ and $10^{42}$--$10^{43}$~\ergs;
$z=$0.5--1, 1--2 and 2--3).  Also included in this plot is the
mean-average \lir\ vs. mean-average \lx\ of the low redshift
BAT/\IRAS\ AGNs.  These averages reproduce the same overall trends as
those seen in the FIR detected fraction of our sample, i.e., that,
within each redshift bin there is no strong correlation between the
average \lir\ and average \lx, but that the average \lir\ increases
strongly with increasing redshift.  The average properties, including
the average \lir, of the AGNs in each of our \lx\ and redshift bins
are presented in \tab{TableAvg}.

The increase in the average \lir\ of our AGNs with increasing redshift
is clearly seen in the right-hand panel of \fig{LIRz}, where we plot
the average \lir\ as a function of redshift.  The mean \lir\ of
\lx=$10^{42}$--$10^{43}$~\ergs\ AGNs increases by a factor of
$3.5^{+1.9}_{-1.2}$
\unskip from $z=0.5-1$ to $z=1-2$ and by
a factor of $10.3^{+7.5}_{-5.0}$
\unskip from $z=0.5-1$ to
$z=2-3$.  The large uncertainty on the latter of these ratios is due
to the small number of \lx=$10^{42}$--$10^{43}$~\ergs\ AGNs in our
highest redshift bin (i.e., 29
\unskip,
of which only 11
\unskip are
detected at FIR wavelengths).  Comparison with the
\lx=$10^{42}$--$10^{43}$~\ergs\ AGNs in the $z<0.1$ BAT/\IRAS\ sample
(mean \lir$=$$2.23^{+0.78}_{-0.58}\times10^{10}$~\lsun
\unskip) reveals an increase in the
mean \lir\ by factors of $5.3^{+3.5}_{-2.3}$
\unskip,
$19^{+11}_{-8}$
\unskip and
$55^{+44}_{-29}$
\unskip from $z<0.1$ to $z=0.5-1$,
$z=1-2$ and $z=2-3$, respectively.\footnote{The errors on the ratios
  presented in this section reflect the maximum and minimum possible
  values, the statistical errors will be roughly 30-40 per cent
  smaller.}  

\begin{figure*}
\begin{center}
	\includegraphics[width=17.0cm]{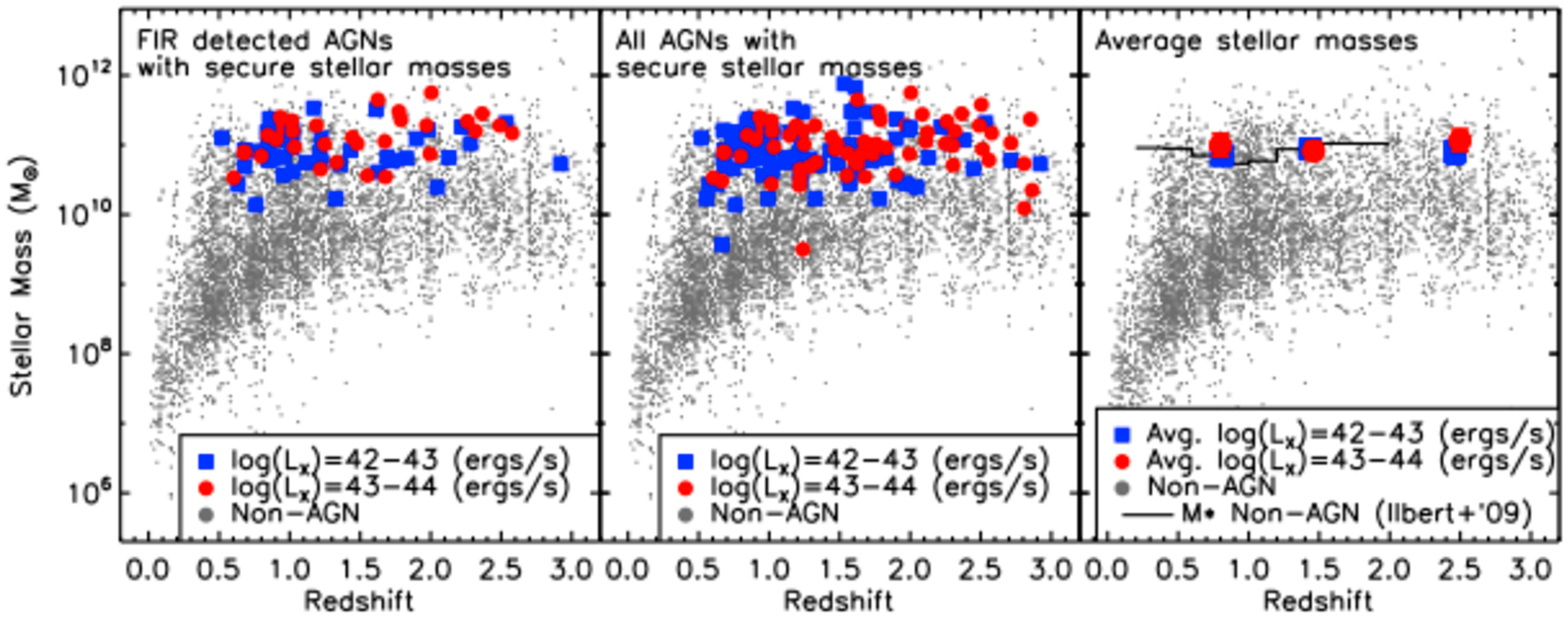}
\end{center}
\caption{{\it Left:} Host galaxy stellar masses (i.e., \mgal)
  vs. redshift for the FIR-detected X-ray AGNs in the \Chandra\ Deep
  fields, split in terms of intrinsic rest-frame 2--10~keV luminosity
  (i.e., \lx; see key).  Stellar masses are calculated by fitting the
  optical--NIR fluxes with the stellar templates of \protect
  \cite{2003MNRAS.344.1000B}.  We have ensured that the light in these
  bands is dominated by the host galaxies, rather than the AGN (see
  \S\ref{Calculating:Masses} and \fig{V_min_J}). The distribution of
  host galaxy masses for AGNs of \lx$=10^{42}$--$10^{44}$~\ergs\ is
  remarkably flat across the considered redshift range. {\it Middle}:
  Same as {\it left} panel, but for all host galaxies for which we can
  measure reliable stellar masses (see
  \S\ref{Calculating:Masses}). {\it Right}: Same as {\it left}, but
  now we show the mean galaxy masses in the $z=0.5-1$, $1-2$ and $2-3$
  bins (separated into \lx\ bins; see key).  Again, there is no clear
  evidence of any significant evolution in the stellar masses of AGN
  host galaxies out to $z\lesssim3$.  In all plots we also show the
  distribution of masses for the underlying galaxy population
  (calculated in the same way as for the AGN hosts) and in the
  right-hand plot we include the values of $M^{\ast}$ (i.e., the point
  at which the galaxy mass function turns over) for star-forming
  galaxies as a function of $z$ as reported in \protect
  \cite{2010ApJ...709..644I}.}
\label{Massz}
\end{figure*}

We see a similar increase in \lir\ with redshift for the more X-ray
luminous AGNs in our sample (i.e., \lx$=10^{43}$--$10^{44}$~\ergs).
Considering just the high redshift (i.e., CDF) sample first, there is
a factor of $3.5^{+1.8}_{-1.2}$
\unskip increase in the mean
\lir\ between the $z=0.5-1$ and $z=1-2$ bins and a factor
$11.3^{+7.0}_{-4.5}$
\unskip increase between the $z=0.5-1$
and $z=2-3$ bins.  Comparing with the mean \lir\ of BAT/\IRAS\ sample
AGNs from the same \lx\ interval (\lir$=$$4.5^{+3.3}_{-1.9}\times10^{10}$~\lsun
\unskip)
reveals an increase in \lir\ by factors of {\it at least}
$2.9^{+3.2}_{-1.6}$
\unskip,
$10^{+11}_{-5}$
\unskip and
$33^{+39}_{-19}$
\unskip from $z<0.1$ to $z=0.5-1$,
$z=1-2$ and $z=2-3$, respectively.\footnote{In reality these ratios
  are likely considerably higher since the majority of
  \lx$=10^{43}$--$10^{44}$~\ergs\ AGNs from the BAT/\IRAS\ comparison
  sample have only upper limits for \lir.} A similar increase in the
\lir\ of \lx$=10^{43}$--$10^{44}$~\ergs\ AGNs with redshift was
reported by \cite{2010ApJ...712.1287L} and \cite{2010A&A...518L..26S}.
These observed increases in \lir\ are also consistent with the upper
limits reported in \cite{2010MNRAS.401..995M} that were derived from
\Spitzer\ 70~\mum\ observations.

As a check, we also calculated the mean \lir\ of AGNs in the full
\lx$=10^{42}$--$10^{44}$~\ergs\ bin and find that they are consistent
with the results from the narrower \lx\ bins (see \fig{LIRz} and
\tab{TableAvg}).  A least-squares fit to the average \lir\ of these
AGNs gives:
\begin{equation}
\label{LIR_t}
L_{\rm IR}/{\rm L_\odot} = 1.4\times10^{13}~\left(t/{\rm Gyr}\right)^{-2.4}
\end{equation}
\noindent
where $t$ is the age of the Universe in Gyr.  The influence of the
FIR-undetected AGNs on these average \lir\ values is clearly seen in
the $z=2-3$ bin, which lies close to the bottom of the distribution of
FIR-detected non-AGNs in the same field.  This explains why the
average \lir\ of our sample does not rise as quickly with redshift as
the FIR-detected galaxies.  

Finally, we note that if we instead use volume-averaged bins,
($=<L_{\rm IR}>N_{\rm AGN}/V(z_1, z_2)$, where $N_{AGN}$ is the number
of AGNs in each redshift bin and $V(z_1, z_2)$ is the surveyed volume
between the upper and lower redshift bounds, $z_1$ and $z_2$; e.g.,
\citealt{Heckman04}) we find that the infrared luminosity density of
\lx$=10^{42}-10^{44}$~\ergs\ AGNs increases by a factor of $\sim$3
from z=0.5-1 to z=2-3. This is in general agreement with the increase
in the total SFR density over the same redshift range (e.g.,
\citealt{Murphy09, Magnelli11}) and implies that the relative galaxy
to black hole growth in moderate luminosity AGNs was higher at high
redshifts compared to today (assuming that the bolometric corrections,
accretion histories etc. have, on average, remained constant
throughout; see also \S\ref{Discussion:Hosts}).

We have shown that the GOODS-H observations confirm the results from
previous studies of the (far) infrared properties of X-ray selected
AGNs over large redshift ranges; i.e., that, within the
\lx$=10^{42}$--$10^{44}$~\ergs\ interval there is no significant
\lx--\lir\ correlation but \lir\, and thus SFRs, increases strongly
with redshift.  In the following subsection we explore whether this
increase in \lir\ with redshift is caused by changes in the stellar
mass or the SSFRs (or a mixture of both) of the AGN hosts. We find
that it is most likely due to the latter and is broadly consistent
with the increase in the SSFRs of star-forming galaxies with
increasing redshift.

\subsection{Masses and specific star formation rates of AGN-hosting
  galaxies.}
\label{Results:Masses}

In their study of the sub-mm properties of X-ray selected AGNs,
\cite{2010ApJ...712.1287L} showed that the increase in the sub-mm
luminosities of \lx$=10^{42}$--$10^{44}$~\ergs\ AGNs is consistent
with the increase in infrared output of normal, star-forming
(hereafter, main-sequence) galaxies.  They argue that this could imply
that moderate luminosity AGNs typically reside in main-sequence
galaxies undergoing internally-induced evolution (as opposed to
major-mergers).  However, there remains other possible explanations
for the increase in \lir\ of these AGNs; changes in the average mass
of AGN-hosting galaxies or a transition from starbursting to
main-sequence hosts could also produce similar trends.  In this
section, we will combine the results outlined above with mass
estimates of the host galaxies (see \S\ref{Calculating:Masses}) to
discriminate between these three scenarios.

\begin{figure}
\begin{center}
	\includegraphics[width=8.2cm]{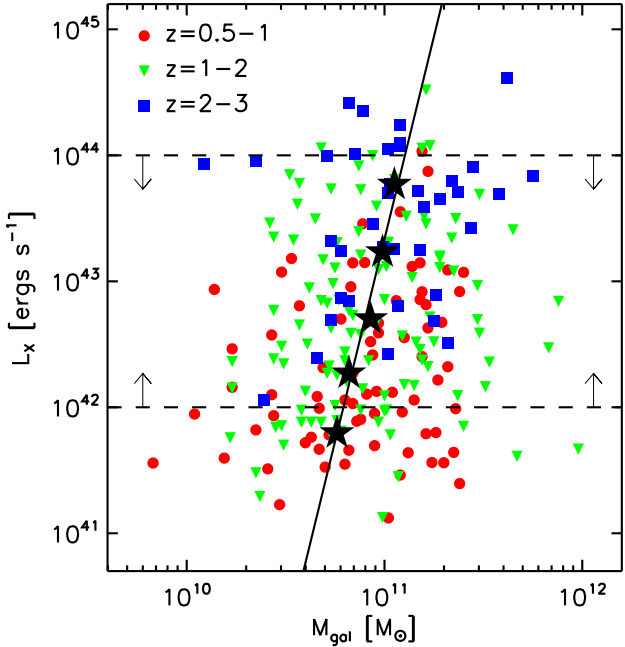}
\end{center}
\caption{Intrinsic (absorption-corrected) rest-frame 2-10~keV X-ray
  luminosity (i.e., \lx) vs. host galaxy stellar mass (i.e., \mgal)
  for the X-ray AGNs in the \Chandra\ Deep Fields, separated in terms
  of redshift (see key).  The dashed lines indicate the
  \lx$=10^{42}$--$10^{44}$~\ergs\ range used throughout this
  study. Large black stars are used to show the median stellar masses
  of AGNs in six different \lx\ bins (0.5 dex wide, starting at ${\rm
    log}\left(L_{\rm X}/{\rm ergs~s^{-1}}\right)=41.5$.  The solid
  line is a least-squares linear fit to these six points.  We find
  only a weak correlation between \lx\ and host stellar masses for the
  AGNs in our sample; over two orders of magnitude increase in \lx\
  (i.e., \lx$=10^{42}$--$10^{44}$) the median stellar mass only
  increases by a factor of $\approx$2.}
\label{LxM}
\end{figure}

To determine whether the increase in \lir\ of moderate luminosity AGNs
is due to an increase in the average host galaxy mass towards higher
redshifts, we plot the stellar masses of the AGNs in our sample as a
function of redshift (see \fig{Massz}).  In the left-hand panel of
\fig{Massz} we plot those \lx$=10^{42}$--$10^{44}$~\ergs\ AGNs that
are detected at FIR wavelengths on top of the underlying distribution
of stellar masses of non-AGNs in the GOODS fields. We find no
significant evolution in the stellar masses of FIR-detected AGN hosts
over the redshift range considered (i.e., $0.5<z<3$), with these
moderate luminosity AGNs residing almost exclusively in
$10^{10}-10^{12}$~\msun\ galaxies at all redshifts
($<$\mgal$>=$$1.02^{+0.17}_{-0.15}\times10^{11}$~\msun
\unskip for
FIR-detected, \lx$=10^{42}$--$10^{44}$~\ergs\ AGNs).
\cite{2010ApJ...720..368X} reported similar findings for
\lx$\sim10^{42}$--$10^{44}$~\ergs\ AGNs in the CDFs out to $z\sim4$
and this mass range is also similar to that of optical narrow-line
selected AGNs in the local Universe (i.e., $0.02<z<0.3$;
\citealt{2003MNRAS.346.1055K}).  We get marginally smaller mean
stellar masses when we consider all AGNs for which we can calculate
secure host galaxy masses (i.e., $\approx$69
\unskip
per cent; see right-panel of \fig{Massz};
$<$\mgal$>=$$8.6^{+1.2}_{-1.0}\times10^{10}$~\msun
\unskip), indicating
that detection at FIR wavelengths biases in favour of slightly higher
host galaxy stellar masses (although we note that the mean masses are
consistent to within 1\sig\ uncertainties).  Interestingly, the AGNs
in our sample show a weak correlation between \lx\ and host galaxy
stellar mass, which can be expressed as:

\begin{equation}
\frac{M_{\rm star}}{\rm M_\odot} \sim 6\times10^4\left(\frac{L_{\rm X}}{\rm ergs~s^{-1}}\right)^{1/7}
\end{equation}
\noindent
This corresponds to a change in the median stellar mass of only a
factor of $\approx2$ across two orders of magnitude change in \lx\
(i.e., \lx$=10^{42}$--$10^{44}$~\ergs; see \fig{LxM}) and is too weak
to account for the prevalence of AGNs among only high mass
galaxies.\footnote{We note that this correlation remains largely
  unchanged when we exclude AGNs from our highest redshift bin (i.e.,
  $z=2-3$), suggesting that selection biases do not have a major
  effect.} In other words, the high host galaxy stellar masses of our
AGN sample is not the result of a selection effect.  We stress that
these results only apply to AGNs with \lx$=10^{42}$--$10^{44}$~\ergs,
it is entirely possible that lower luminosity AGNs (including those
below the detection threshold of the CDF surveys) reside in lower mass
galaxies.  However, since there is no significant evolution in the
host galaxy stellar masses of moderate luminosity AGNs we can rule out
the possibility that the observed increase in their average \lir\ is
due to a systematic increase in their host galaxy stellar mass with
redshift.  Instead, we must attribute it to an increase in their
SSFRs.  Next, we consider whether this increase in their SSFRs is
consistent with that observed in the general star-forming (i.e.,
main-sequence) galaxy population, or whether it signifies a
transitioning of AGN host galaxies from starbursting galaxies at
$z=2-3$ to main-sequence/quiescent galaxies today.

In the formalism of \cite{Elbaz:2011uw}, secure starburst galaxies
have SSFRs that are $\gtrsim 3$ times higher than those of
main-sequence galaxies (here, ``secure'' implies avoiding any
contamination from main-sequence galaxies).  This is the case at all
redshifts, although the SSFRs of main-sequence galaxies increase at
higher redshifts.  Thus, to determine whether the observed \lir--$z$
trend for X-ray AGNs is caused by a transitioning of their hosts from
starbursting to main-sequence galaxies, we will compare their SSFRs
over the $0.5<z<3$ redshift interval of our sample.

\begin{figure*}
\begin{center}
	\includegraphics[width=16.0cm]{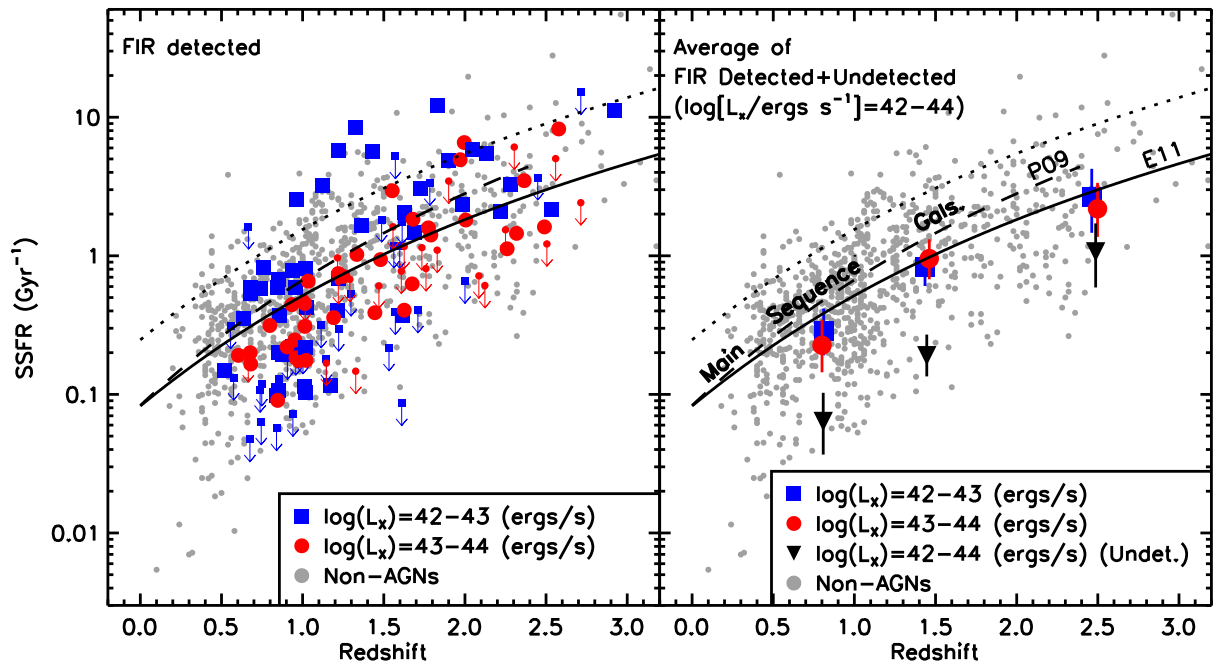}
\end{center}
\caption{{\it Left}: The specific star formation rates (i.e., SSFRs)
  of the host galaxies of the X-ray AGNs in our sample, plotted
  against redshift and separated in terms of intrinsic rest-frame
  2-10~keV luminosity (see key). Downward pointing arrows are used to
  indicate the 3\sig\ upper limits on SSFR in cases where there is no
  formal FIR detection. The vast majority of FIR-detected AGN reside
  in normal, star-forming (i.e., main-sequence) galaxies. {\it Right}:
  Same as the left panel, but for the mean SSFRs calculated using the
  mean star formation rates (i.e., SFRs) and stellar masses (i.e.,
  \mgal) of the AGNs in various redshift and \lx\ bins (see key).
  Also included in this plot are the average SSFRs calculated from the
  stacks of the FIR-undetected AGNs which we use to estimate the
  fraction of AGNs that reside in quiescent galaxies (solid black
  triangles; see \S\ref{Results:Masses}).  For comparison, we include
  in both panels the non-AGNs detected in the GOODS-H survey
  (restricted to \mgal$=10^{10}-10^{12}$~\msun\ to match the range of
  stellar masses spanned by our AGN sample; see \S\ref{Results:Masses}
  and \fig{Massz}).  Also included are tracks showing the SSFRs of
  normal, star-forming (i.e., main-sequence) galaxies at the redshifts
  considered (shown by the solid and dashed lines; from \protect
  \citealt{Elbaz:2011uw} and \protect \citealt{2009ApJ...698L.116P},
  respectively). The dotted line is a factor of 3 above the \protect
  \cite{Elbaz:2011uw} curve, above which a galaxy can be safely
  considered to be strongly starbursting (see \protect
  \citealt{Elbaz:2011uw}).}
\label{SSFRz}
\end{figure*}

In \fig{SSFRz} we plot the SSFRs of the
\lx$=10^{42}$--$10^{44}$~\ergs\ AGN-hosting galaxies in our sample as
a function of redshift. Included in these plots are the SSFRs of
normal (i.e., non-AGN-hosting) galaxies with
\mgal$=10^{10}$--$10^{12}$~\msun\ detected in GOODS-H, i.e., only
those galaxies with stellar masses within the range of the AGNs in our
sample.  We also include tracks showing the SSFR-$z$ trends for
main-sequence galaxies reported in \cite{2009ApJ...698L.116P} and
\citeauthor{Elbaz:2011uw} (\citeyear{Elbaz:2011uw}; see their
Eq. 13).\footnote{\label{ElbazFit} A fit to the SSFRs of the
  comparison \mgal$=10^{10}$--$10^{12}$~\msun\ galaxies shown in
  \fig{SSFRz} reproduces Eq. 13 of \cite{Elbaz:2011uw}.} In the left
hand panel of \fig{SSFRz} we plot the SSFRs for the FIR-detected AGNs
in our sample.  The majority (i.e.,
$\approx$81 per cent
\unskip) of these AGNs lie within a
factor of 3 of the \cite{Elbaz:2011uw} main-sequence trend, with
roughly equal numbers of FIR-detected AGNs outside this range having
higher or lower SSFRs (i.e., $\approx$47 per cent
\unskip
and $\approx$53 per cent
\unskip of outliers,
respectively).  However, none of the FIR-detected AGNs in our sample
have SSFRs lower than 17 per cent
\unskip of the
\cite{Elbaz:2011uw} main-sequence and are thus unlikely to be
quiescent galaxies (c.f., \citealt{2007ApJ...660L..43N}).  Later in
this section we use the results from our stacking analyses to estimate
what fraction of {\it all} (i.e., FIR detected and undetected)
\lx$=10^{42}$--$10^{44}$~\ergs\ AGNs reside in quiescent galaxies.
Over the entire $0.5<z<3$ redshift range only
 8
\unskip (i.e.,
$\approx$10 per cent
\unskip) of the
77
\unskip FIR-detected,
\lx$=10^{42}$--$10^{44}$~\ergs\ AGNs with secure host galaxy stellar
masses have SSFRs typical of starburst galaxies at their redshifts.
This number rises to 11
\unskip when
we include the 3\sig\ upper limits of FIR-undetected sources (i.e.,
$\approx$ 8 per cent
\unskip of all
130
\unskip AGNs within these redshift and \lx\
bounds).

The trends seen in the FIR-detected fraction of our AGN sample are
reflected in the average SSFRs derived from our stacking analyses (see
\fig{SSFRz}, right-hand panel).  The average SSFRs of the AGN host
galaxies in our sample lie slightly below the \cite{Elbaz:2011uw}
trend for main-sequence galaxies. At $z=0.5-1$, $1-2$ and $2-3$,
\lx$=10^{42}$--$10^{44}$ AGNs have average SSFRs that are
$71^{+24}_{-19}$
\unskip,
$89^{+21}_{-19}$
\unskip and
$72^{+28}_{-21}$
\unskip per cent of those of typical
main-sequence galaxies at their average redshifts, respectively (i.e.,
$\overline{z}=$$0.81$
\unskip,
$1.4$
\unskip and
$2.5$
\unskip).\footnote{The corresponding results
  for the \lx$=10^{42}$--$10^{43}$~\ergs\ bins are
  $75^{+34}_{-25}$
\unskip,
  $87^{+28}_{-23}$
\unskip and
  $91^{+55}_{-41}$
\unskip per cent; for the
  \lx$=10^{43}$--$10^{44}$~\ergs\ bins they are
  $60^{+32}_{-22}$
\unskip,
  $98^{+37}_{-28}$
\unskip and
  $73^{+39}_{-27}$
\unskip per cent.}  Since we
are effectively comparing against the average SSFRs of
\mgal$=10^{10}$--$10^{12}$~\msun\ galaxies (i.e., see footnote
\ref{ElbazFit}), it is unlikely that these lower SSFRs are due to the
AGN hosts having higher-than-average stellar masses.  Despite being
slightly lower, the mean SSFRs of the AGNs in our sample show largely
the same magnitude increase as main-sequence galaxies, irrespective of
\lx; a factor of $3.1^{+2.1}_{-1.3}$
\unskip from
$\overline{z}=$$0.81$
\unskip to
$1.4$
\unskip and a factor of
$7.8^{+6.9}_{-3.7}$
\unskip from
$\overline{z}=$$0.81$
\unskip to
$2.5$
\unskip for \lx$=10^{42}$--$10^{44}$~\ergs\
AGNs.

As we have already seen, none of the FIR-detected AGNs in our sample
have SSFRs typical of quiescent galaxies (i.e., $<10$ per cent of the
average SSFR of main-sequence galaxies at a given redshift).  However,
this still leaves the $\approx$60 per cent of AGNs in our sample that
remain undetected at FIR wavelengths.  We can use the results from our
stacking analyses to place constraints on the fraction of these
FIR-undetected, \lx$=10^{42}$--$10^{44}$~\ergs\ AGNs that reside in
quiescent galaxies.  Adopting a Monte Carlo approach, we randomly
assign a $<3\sigma$ flux to each FIR-undetected source assuming an
underlying log-normal SSFR distribution (Rodighiero et al. in prep.),
then calculate the average SSFR of these simulated sources. We repeat
this process 1000 times, selecting only those trials whose simulated
average SSFRs lie within the minimum and maximum bounds derived from
our stack.  For each successful trial, we calculate the fraction of
AGNs with simulated SSFRs below 10 per cent of SSFR$_{\rm MS}$ (i.e.,
quiescent).  In doing so, we estimate that between 13 and 37 (i.e.,
$\approx$16 and $\approx$46 per cent, respectively) of the 81
FIR-undetected AGNs between $z=0.5-3$ with securely-determined stellar
masses have SSFRs that are consistent with those of quiescent
galaxies.  These numbers correspond to $15\pm7$ per cent of all
\lx$=10^{42}$--$10^{44}$~\ergs\ AGNs within this redshift interval.
This fraction is similar in size to the average deficit of AGN SSFRs
compared to the main-sequence trend, suggesting that the inclusion of
quiescent galaxies in our average SSFRs is responsible for this offset
(see also \S\ref{Discussion:Impact}); \footnote{By excluding the
  quiescent 15$\pm$7 per cent of hosts from our average SSFRs, we
  estimate that the SSFRs of star-forming AGN hosts are $\sim$95 per
  cent of those of typical main-sequence (i.e., non-AGN) galaxies (and
  consistent to within the various uncertainties).}.  By incorporating
the results derived from the FIR-detected AGNs, we estimate that
$7\pm2$ per cent of \lx$=10^{42}$--$10^{44}$ AGNs at $z=$0.5--3 reside
in starbursting galaxies and $79\pm10$ per cent reside in
main-sequence galaxies.  These fractions do not change significantly
with redshift and show a slight preference toward main-sequence
galaxies compared to the general population.  For example,
\cite{2007ApJ...660L..43N} found that $\approx$56--67 per cent of
$0.2<z<1.1$ galaxies (with similar stellar masses as the AGN hosts
considered here) are main-sequence, while $\approx$30 per cent are
quiescent and $\approx$1 per cent are starbursts (although
\cite{2007ApJ...660L..43N} defined starbursts as having SSFRs $>$5
times higher than that of main sequence galaxies, rather than $>$3
times used here).

Whilst we can only place upper limits on the fractions of AGNs that
reside in quiescent and main-sequence galaxies, we are able to report
that the SSFRs outlined here are not compatible with a view that AGN
hosts tend to be strongly starbursting galaxies, nor that moderate
luminosity AGNs have transitioned from residing in starburst galaxies
at high redshifts to main-sequence galaxies at $z\sim0.5$.  This is in
general agreement with \cite{2010ApJ...720..368X}, although they found
that $z<1$ X-ray AGNs in the CDFs have SFRs that are a factor of 2-3
times higher compared to a stellar mass-matched sample of non-AGNs.
However, we note that their SFRs were derived from 24~\mum\
observations, which are more susceptible to contamination from either
emission from the AGN or strong spectral features (i.e., PAH lines).

Our results disfavour changes in either the stellar masses of AGN
hosts and/or a transitioning of their hosts from
main-sequence/quiescent to starbursting galaxies as giving rise to the
observed increase in AGN \lir\ at higher redshifts.  We have also
shown that the SSFRs of AGN hosts are only marginally lower (i.e.,
$\approx20$ per cent lower), than those of main-sequence galaxies at
$0.5<z<3$.  This extends the work of \cite{2010ApJ...712.1287L} by
quantifying their suggestion that moderate luminosity AGNs typically
reside in main-sequence galaxies undergoing internally-induced
evolution.  We further explore the consequences of this interpretation
in \S\ref{Discussion}.

\subsubsection{The red rest-frame colours of AGN-hosting galaxies}
\label{Results:RedColours}

\begin{figure*}
\begin{center}
	\includegraphics[width=14.0cm]{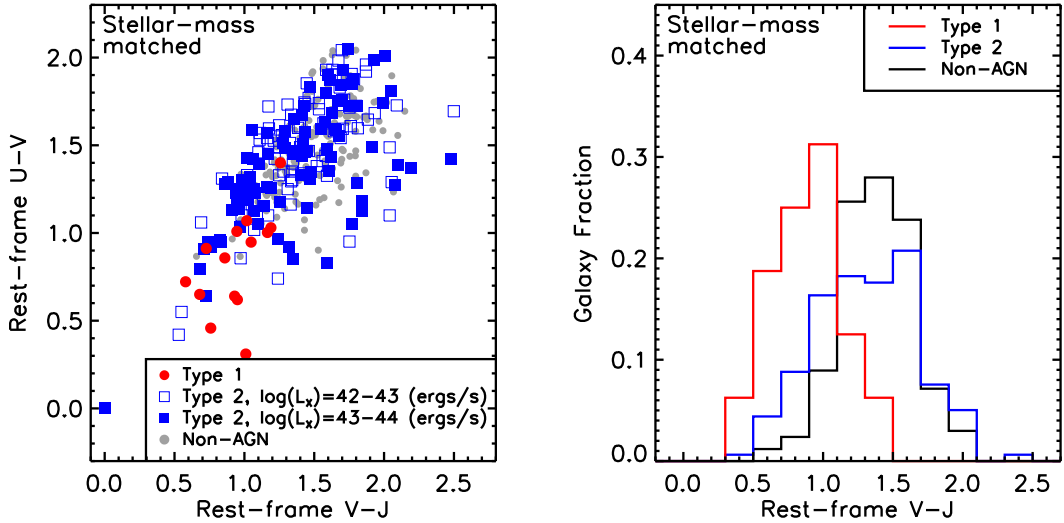}
\end{center}
\caption{Same as \fig{V_min_J}, but now we compare the colours of our
  sample of X-ray AGNs against only \Herschel-detected main-sequence
  galaxies with stellar masses (i.e., \mgal) within $\pm$0.2 dex of
  the characteristic \mgal\ of AGN host galaxies (i.e.,
  $1.53\times10^{11}$~\msun).  The host galaxy colours of moderate
  luminosity (i.e., \lx$=10^{42}$--$10^{44}$~\ergs\ AGNs is remarkably
  similar to those of main-sequence galaxies with similar stellar
  masses (see \S\ref{Results:RedColours}).}
\label{V_min_J_Herschel}
\end{figure*}

\begin{figure}
\begin{center}
	\includegraphics[width=8.0cm]{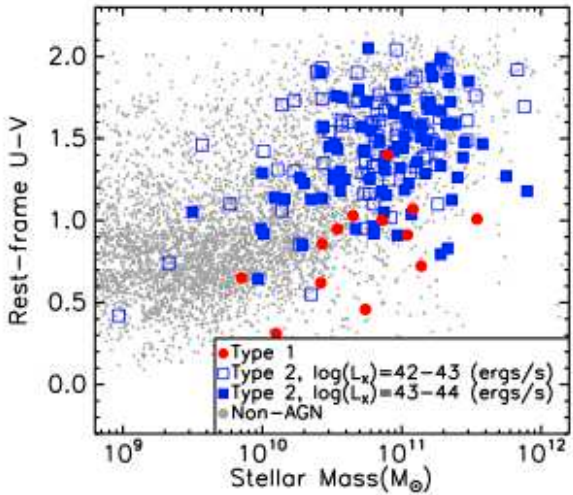}
\end{center}
\caption{Plot of rest frame $U-V$ vs. host galaxy stellar mass for the
  AGNs in our sample (large symbols) and non-AGNs in the same fields
  (small symbols).  The rest frame colours have not been corrected for
  the effects of reddening.  There is a tendency for the AGNs in our
  sample tend to reside in either the so-called ``red sequence'' or
  ``green valley'', despite their far-infrared derived SFRs and SSFRs
  indicating that the majority are, in fact, star-forming (i.e.,
  main-sequence) galaxies.  We conclude that the reddened colours of
  these massive, star-forming AGN hosts are largely due to the effects
  of dust obscuration.  Here, we have included all \Herschel\ detected
  AGNs, not just those with ``secure'' masses (see
  \S\ref{Calculating:Masses}), so as not to bias the plot.}
\label{U_min_V_Mass}
\end{figure}

Here we take a brief aside to address a specific question regarding
the rest-frame optical colours of the AGN host galaxies explored in
this study: if the majority of moderate-luminosity AGNs reside in
star-forming, main-sequence galaxies, then why are their host galaxies
generally redder than usually expected for these types of galaxies
(i.e, see \fig{V_min_J})?  The $\sim20$ per cent difference in the
average SSFRs of AGN hosts is too small to account for their red
colours.  Instead, the answer likely lies in the high masses of AGN
host galaxies.  More massive star-forming galaxies tend to be redder
than their lower mass counterparts with similar SSFRs due to increased
levels of dust obscuration in these galaxies (e.g.,
\citealt{2009ApJ...698L.116P}).  Indeed, when we compare the
rest-frame colours of our (\lx$=10^{42}$--$10^{44}$~\ergs) AGN hosts
against those of non-AGN-hosting, main-sequence galaxies with similar
stellar masses (i.e., within $\pm$0.2 dex of the median stellar mass
of the AGN hosts), we find that they are remarkably similar (see
\citealt{2010ApJ...720..368X} and our \fig{V_min_J_Herschel}).  This
is contrary to a number of recent studies reporting that X-ray AGNs
often have rest-frame colours offset from the majority of the galaxy
population (i.e., within the so-called ``green valley'', rather than
either the ``red sequence'' or ``blue cloud''; see our
\fig{U_min_V_Mass}), which has been interpreted as evidence of the
quenching of star formation by the AGN
(e.g. \citealt{2007ApJ...660L..11N, 2007MNRAS.382.1415S,
  2008MNRAS.385.2049G, 2008ApJ...681..931B, 2008ApJ...675.1025S}).
Instead, once the effects of reddening are correctly taken into
account, AGN hosts display largely the same distribution of intrinsic
colours as the general galaxy population, as was recently shown by
\cite{2010ApJ...721L..38C}.  As further confirmation of this, the
fractions of main-sequence and quiescent AGN hosts reported here (see
\S\ref{Results:Masses}) are broadly consistent with the fractions of
star-forming (blue cloud) hosts reported in
\cite{2010ApJ...721L..38C}, although they report slightly higher
fractions of quiescent (red sequence) hosts (likely due to the
different approaches used to identify quiescent galaxies, i.e.,
infrared vs. optical diagnostics).  Thus, the hosts of moderate AGNs
have largely the same fundamental stellar properties (i.e., rest-frame
colours, SFRs, SSFRs) as their non-AGN-hosting counterparts of similar
stellar mass.

\subsection{The optical to far-infrared SEDs of X-ray AGNs}
\label{Results:SEDs}

\begin{figure}
\begin{center}
	\includegraphics[width=8.0cm]{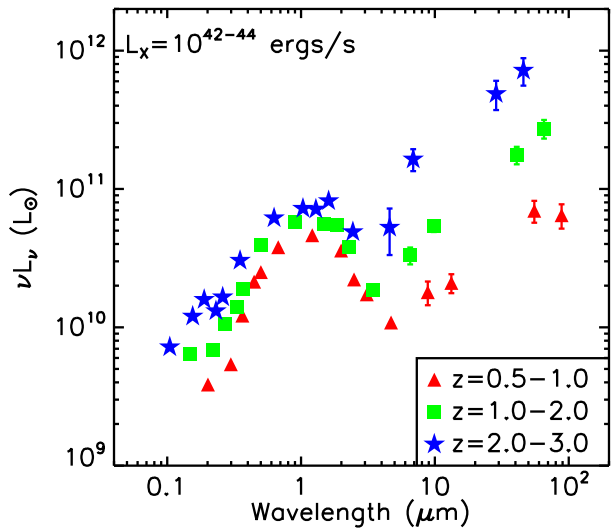}
\end{center}
\caption{Average optical to far-infrared SEDs of the
  \lx$=10^{42}-10^{44}$~\ergs\ AGNs in our three moderate/high
  redshift bins (see key).  The strong increase in the average
  infrared luminosities of the AGNs is clearly evident in these SEDs.
  A strong stellar component is evident in all three SEDs around
  1~\mum, demonstrating that, on average, the host galaxy dominates
  the optical and near-infrared fluxes.  As further confirmation of
  this, there is even evidence of the 1.6~\mum\ stellar ``bump'' in
  the average SED of the $z=2-3$ AGNs (where the mean redshift of the
  AGNs places this feature within the IRAC 5.6~\mum\ filter).}
\label{OptSED}
\end{figure}

If the observed increase in \lir\ of AGNs with redshift is solely due
to the evolving SSFR of their main-sequence hosts, as our results seem
to suggest, we should expect their infrared SEDs to be increasingly
dominated by star formation at higher redshifts. In \fig{OptSED} we
plot the average rest-frame SEDs traced out by the optical to
far-infrared bands for \lx$=10^{42}-10^{44}$~\ergs\ AGNs in each of
our redshift bins.  While the optical to near infrared portions of the
average AGN SEDs show only modest change with redshift, both the mid
(i.e., 16 and 24~\mum) and far-infrared (i.e., 100 and 160~\mum)
portions show a strong increase with increasing redshift, as
previously reported by \cite{2010MNRAS.401..995M}.  This increase is
strongest at FIR wavelengths meaning that the average infrared SED of
\lx$=10^{42}$--$10^{44}$~\ergs\ AGNs is redder at higher redshifts.

\begin{figure*}
\begin{center}
	\includegraphics[width=14.0cm, height=8.2cm]{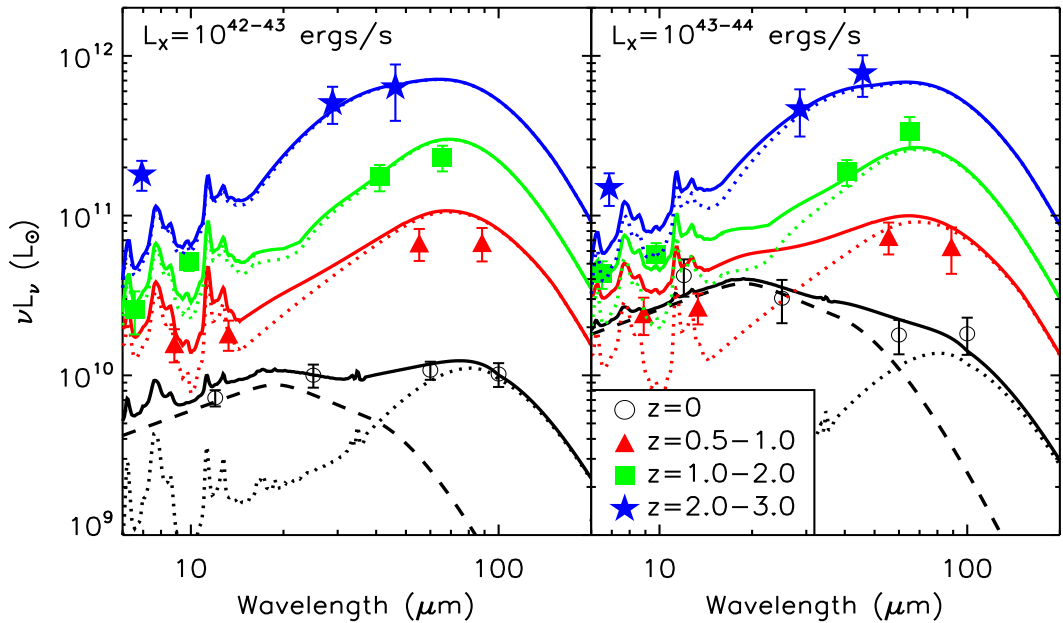}
\end{center}
\caption{The average observed infrared spectral energy distributions
  (SEDs) of the AGNs in the \Chandra\ Deep Fields and the local
  BAT/\IRAS\ sample (points).  Again, we have separated our sample in
  terms of X-ray luminosity (left and right panels; see labels).  Also
  shown on these plots is the expected SEDs calculated by boosting the
  host galaxy component of the fit to the $z\approx0$ observed SED by
  a factor equivalent to the increase in the SSFR at the average
  redshift of each bin (solid lines, colour corresponding to the
  appropriate redshift bin; see \S\ref{Results:SEDs} for a full
  description).  The dashed black line indicates the AGN component at
  all redshifts, while the dotted lines (various colours) represent
  the host galaxy component at the mean redshift of our subsamples
  (taken from the \protect \citealt{2001ApJ...556..562C} library).  It
  is important to note that the solid lines are {\it not} an attempted
  fit to the data, but the construct of a model based on the simple
  assumption that only the normalisation of the host-galaxy component
  increases with redshift.  This plot demonstrates that the increase
  in luminosity observed in each infrared band (i.e., 16~\mum,
  24~\mum, 100~\mum\ and 160~\mum) is roughly consistent with the
  increase in SSFR observed in normal, star-forming galaxies (see
  \S\ref{Results:SEDs}).}
\label{CompareSEDs}
\end{figure*}

Is the change in the average AGN SED consistent with pure evolution of
the SSFR of main-sequence galaxies? To assess this we predict the
infrared SEDs of high redshift AGNs by crudely assuming that {\it
  only} the SSFRs of their host galaxies have evolved with redshift,
then compare the observed SEDs against these predictions. To generate
the predicted SEDs we first fit the average infrared SED of $z\sim0$
BAT/\IRAS\ AGNs with a model comprising an AGN and a single host
galaxy component (see \citealt{2011MNRAS.414.1082M} for a description
of the fitting procedure).  The AGN component derived from this fit is
then kept constant throughout.  To generate the host-galaxy components
for the high redshift AGNs, we estimate the expected \lir\ of the
hosts based on their average stellar masses and redshifts (using Eq. 2
of \citealt{2009ApJ...698L.116P}).  We then select the SED from the
\cite{2001ApJ...556..562C} library that most closely matches this
\lir\ and add it to the AGN component from the $z\sim0$ fit to produce
the total predicted SED (i.e., AGN+host).  The SEDs generated in this
way are shown in \fig{CompareSEDs} and are roughly consistent with the
observed mean AGN SEDs in each redshift bin (the slight discrepancy at
MIR wavelengths for our $z=2-3$ bin is likely due to the increased
strength of PAH features in high-z star-forming galaxies compared to
the local templates used here; \citealt{Chary10, Elbaz:2011uw}).
Since these predictions only take into account the {\it change} in the
SSFR from $z\sim0$, we may not expect to see any evidence of the
reduced SSFRs of AGNs compared to main-sequence AGNs (see
\S\ref{Results:LIR:Undetected}) in this plot, especially if $z\sim0$
AGNs also have similarly reduced SSFRs (e.g.,
\citealt{2007ApJS..173..267S}).  Whilst this is only a crude analysis,
it does show that the whole AGN+host galaxy infrared SED, rather than
just the FIR portion, evolves in a manner broadly consistent with an
evolving main-sequence SSFR.  An important prediction that arises from
this scenario is that the average MIR spectra of AGNs are increasingly
dominated by host galaxy spectral features (e.g., PAH features) at
higher redshifts.  We will use archival \Spitzer-IRS spectra to
determine whether this is the case in a future paper.

\section{Discussion}
\label{Discussion}

We have used the deepest FIR data currently available to probe the
SFRs and SSFRs of AGN host galaxies to $z\lesssim3$. Our main results
are: (a) there is no correlation between star formation and \Lx\ in
moderate luminosity AGNs (i.e., \lx$=10^{42}$--$10^{44}$~\ergs), (b)
there is a strong, monotonic, increase in the SFRs and SSFRs of the
hosts of these AGNs from $z\sim0$ to 3 (c) that $<10$ per cent of AGNs
reside in starbursting galaxies, the majority being hosted by
main-sequence, star-forming, galaxies and (d) the increase in the
SSFRs of the AGN hosts is broadly consistent with the increase in the
SSFRs of main-sequence galaxies in general. In this section, we
explore the possible implication of these results in terms of our
understanding of the co-evolution of AGNs and their host galaxies.

\subsection{The influence of moderate luminosity AGNs on
  star formation}
\label{Discussion:Impact}

The lack of any strong correlation between \lx\ and \lir\ in any of
our redshift bins suggests that {\it instantaneous} nuclear and global
star formation activity are largely decoupled for moderate luminosity
AGNs at $z<3$.  At first sight, this would seem to imply that nuclear
activity has little impact on the levels of star formation taking
place within their host galaxies, bringing the role of AGN/galaxy
feedback in the evolution of these systems into question.  However, it
has recently been reported that nuclear (i.e., $<1$~kpc), rather than
global, SFRs are strongly correlated with black hole accretion rate
and, thus, AGN luminosity (\citealt{2011arXiv1106.3565D}).  This would
suggest that the physical links between black hole growth and
star-formation activity are localised to central regions of the host
galaxy.  We should also consider the effects that X-ray variability
will have on our results.  Since X-ray observations only take a
snapshot of nuclear activity, the lack of a correlation between \lx\
and \lir\ could simply be due to AGN variability introducing scatter
(e.g., \citealt{2006ApJS..166....1H}). One way to test this, which is
beyond the scope of this paper, would be to use X-ray stacking
analyses to calculate the average \lx\ of {\it all} galaxies as a
function of \lir, assuming that a wide cross-section of the galaxy
population will sample all stages of AGN variability at a given epoch.

The evidence that the SSFRs of galaxies hosting moderate luminosity
AGNs are marginally (i.e., $\approx20$ per cent) lower than their
non-AGN-hosting counterparts, whilst tentative, may hint to a causal
connection between nuclear activity and star formation in their hosts.
However, with the data used here, it is difficult to determine whether
the AGN is playing a direct role in the transitioning of a galaxy from
the main-sequence to a more quiescent state (i.e., quenching).  As we
suggested in \S\ref{Results:Masses}, it may simply be the case that a
non-negligible fraction of moderate luminosity AGNs are triggered in
quiescent galaxies, rather than in star-forming galaxies galaxies
which then transition to a more quiescent state.  Besides, the
difference is so slight (and in some of our redshift bins, negligible)
it is difficult to argue that quenching by moderate levels of nuclear
activity has any significant impact on the star-formation activity of
their host galaxies.

As most previous studies of AGN hosts have largely focussed on SFRs,
rather than SSFRs, it is difficult to directly compare our findings
with these earlier studies.  In any case, they have arrived at
conflicting conclusions, reporting increased (e.g.,
\citealt{2006MNRAS.366..480G}, \citealt{2007ApJ...669L..61L}),
suppressed (e.g., \citealt{2005ApJ...629..680H},
\citealt{2006ApJ...642..702K}) or no significant difference (e.g.,
\citealt{2009ApJ...696..396S}) in the SFRs of AGN hosts depending on
the type/luminosity of the AGNs in their samples and/or selection
method.  By focussing on SFRs (rather than SSFRs) these earlier
studies were susceptible to differences between the masses of their
AGN and control (galaxy) samples. However, by comparing the SFRs of
mass-matched samples of galaxies (synonymous to comparing SSFRs),
\cite{2010ApJ...720..368X} found that the SFRs of AGN hosts at $z=0-1$
are a factor of 2-3 higher than those non-AGNs, whereas this
difference largely disappears at $z=1-3$.  By contrast, in their study
of the ultraviolet-derived SSFRs of galaxies in the local Universe,
\cite{2007ApJS..173..267S} found that AGNs have suppressed SSFRs
compared to star-forming galaxies, although they found that AGNs had,
on average, $\sim60$ per cent lower SSFRs, compared to the $\sim20$
per cent reported here (see their Figs. 17 and 18).  At face value,
this latter point may suggest that nuclear activity plays a stronger
role in quenching star-formation at low redshifts.  However, it is
important to note that our sample selection and analyses differ
considerably from that of \cite{2007ApJS..173..267S}, which may
account for some of the disagreements between these studies.  In the
end, clear evidence of AGN-driven feedback (e.g.,
\citealt{2006ApJ...650..693N, 2010MNRAS.402.2211A}) will be needed to
demonstrate that any suppression of star formation is due to them
having a direct, causal influence on their host galaxies.

\subsection{The majority of moderate luminosity, distant AGNs reside
  in normal, star-forming galaxies at all redshifts.}
\label{Discussion:Hosts}

Our results provide convincing evidence that the majority of distant,
moderate luminosity AGNs reside in main-sequence, star-forming
galaxies. To interpret the implications of this result in terms of the
co-evolution of AGNs and their host galaxies, we will consider the
typical characteristics of galaxies that lie both on and off the
main-sequence.

Main-sequence galaxies tend to be late-type spirals undergoing
internally driven evolution (e.g., \citealt{2007ApJ...660L..43N}).
Star formation in these galaxies predominantly takes place in their
disks and is likely triggered by internal processes such as turbulence
and disk instabilities (e.g., \citealt{2007ApJ...658..763E,
  2008ApJ...687...59G}).  In contrast, starbursting galaxies - the
other main category of star-forming galaxies - often show signs of
having recently undergone a major merger event (mass ratio $\lesssim
3:1$), meaning their star formation is more likely to be merger
induced (e.g., \citealt{2007ApJ...670..156D, 2007A&A...468...33E}).
It is important to note that the starburst phase of a typical major
merger is relatively short lived, with only around $^1/_3$ of the life
of the merger being spent in this regime
(\citealt{2008A&A...492...31D}).  However, for the rest of the time,
the merger is not actively accelerating star formation beyond what
would be expected under normal, internal processes.  Finally, at the
other extreme are quiescent galaxies that have ceased to be actively
forming stars and have SSFRs significantly lower (i.e., $\ll 10$ per
cent; \citealt{2007ApJ...660L..43N}) than those of typical
main-sequence galaxies.

Based on this emerging picture, the results presented here strongly
suggest that moderate levels of nuclear activity have mostly taken
place in galaxies undergoing internally induced, low efficiency star
formation, as opposed to starbursting mergers -- a view consistent
with the conclusions of \cite{2010ApJ...720..368X}.  Of course a
fraction of these AGNs may reside in merging systems that have ceased
starbursting, although even a conservative estimate places this
fraction at around $21\pm6$ per cent.\footnote{Based on $7\pm2$ per
  cent of the AGNs in our sample having starburst-level SSFRs, and
  assuming that a typical merger spends only $^1/_3$ of its time in a
  starburst phase.}  However, even if this is the case, any merging
has since ceased triggering starbursts, meaning the merger event is no
longer having a significant effect on the star formation activity of
these galaxies.  Furthermore, since it seems unlikely that a merger
can channel material onto a black hole without also inducing
significant amounts of star formation, this interpretation also
implies that the majority of moderate nuclear activity has been
triggered and driven by internal processes since $z\sim3$.  Since
main-sequence galaxies at high redshifts (i.e., $z\geq1$) are often
dominated by gravitationally unstable, clumpy disks (e.g.,
\citealt{1996AJ....112..839C, 2004ApJ...604L..21E,
  2007ApJ...658..763E, 2006ApJ...645.1062F, 2009ApJ...706.1364F,
  2008ApJ...687...59G}), we speculate that disk instabilities are
driving the inflows that lead to nuclear activity in their cores.
Indeed, theoretical models have show that the giant clumps can
directly fuel black hole growth (e.g., \citealt{2008ApJ...684..829E}),
and can more generally trigger a central gas inflow fuelling an AGN
(Bournaud \& Dekel, in prep.).

These findings are also consistent with the conclusions of recent
studies that have looked at recent merger histories and morphologies
of galaxies hosting moderate luminosity AGNs.  For example, when they
compared the morphologies of $z\lesssim1$ AGN-hosting galaxies with
those of the general galaxy population, \cite{2011ApJ...726...57C}
showed that the same fraction of each sample showed signs of recent or
ongoing mergers, while \cite{2011ApJ...727L..31S} used a subsample of
28 X-ray AGNs in the CDFS to show that this was also likely the case
at $z\lesssim3$ (see also Kocevski et al, submitted, who have
performed morphological analyses for a larger sample of AGNs, many of
which overlap the sample used in this study).  Our results build on
these earlier findings by using infrared-derived SSFRs - arguably a
more reliable and quantifiable means of identifying main-sequence
galaxies - to show that the vast majority of {\it all} moderate
nuclear activity in these deep fields is accompanied by
internally-induced, rather than violent merger-induced star-formation.

Whilst this interpretation appears to apply for moderate luminosity
AGNs at all redshifts since $z\sim3$, it is particularly pertinent at
$z\lesssim1$, when these AGNs dominated the integrated black hole
growth taking place in the Universe. Indeed, integrating the analytic
AGN X-ray luminosity functions of \cite{2003ApJ...598..886U} and
assuming the luminosity-dependent bolometric correction factor of
\cite{2004MNRAS.351..169M} reveals that
\lx$=10^{42}$--$10^{44}$~\ergs\ AGNs have accounted for $\approx$65
per cent of all black hole growth since $z\sim1$.  Therefore, it would
seem that the majority of the volume-averaged black hole mass
accretion since $z\lesssim1$ has taken place in galaxies undergoing
internally-induced evolution, which also appears to be the case for
the majority of stellar mass build up over this redshift range (e.g.,
\citealt{2005ApJ...625...23B, 2009ApJ...697.1971J}).  What this means
for the concurrent build up of black hole and stellar mass to today's
\mbmb\ relation will be explored in a follow-up paper (Mullaney et
al., in prep.) but it is important to bear in mind that, by assuming
the same luminosity functions and bolometric corrections (although see
footnote \ref{LumDepBol}), we calculate that the majority (i.e.,
$\approx60$ per cent) of {\it all} black hole mass accretion (i.e.,
growth) in the Universe since $z\sim3$ would have taken place in more
luminous AGNs (i.e., \lx$\geq10^{44}$~\ergs).

On the other hand, evidence supporting the importance of mergers on
the concurrent growth of black holes and their host galaxies has been
presented in a number of previous studies.  For example,
\cite{2005Natur.434..738A} showed that $\sim30-50$ per cent of sub-mm
galaxies, which often lie above the main-sequence (in terms of SSFR),
contain moderate to high luminosity (i.e.,
log$[$\lx/\ergs$]\gtrsim43.5$) AGNs (see also, e.g.,
\citealt{2005ApJ...625L..83L, 2007ApJ...660.1060V,
  2007ApJ...655L..65M, 2008ApJ...675.1171P, 2010ApJ...713..503C}).
Furthermore, \cite{2004ApJ...611L..85P} showed that a significant
proportion of luminous, absorbed quasars reside in strongly
star-forming galaxies (although some of these may now be recognised as
main-sequence galaxies at high redshifts).  More recently,
\cite{2011ApJ...729L...4F} identified two Compton-thick, luminous
(i.e., \lx$>10^{44}$~\ergs) AGNs in the CDF-S whose host galaxies are
also undergoing a period of intense star formation.  Taken in
conjunction with the arguments presented above it is tempting to
speculate that, while internal processes have largely driven moderate
nuclear activity, external (i.e., non-secular, major-merger driven)
processes have played a more major, perhaps dominant, role in the
build up of black hole and galaxy bulge mass and, thus, formed today's
\mbmb\ relationship -- as suggested by the theoretical models of
\cite{2006ApJS..166....1H} and recently discussed in
\cite{2010ApJ...712.1287L}.  However, this interpretation would be
difficult to reconcile with emerging evidence that long lived,
internal processes were the main driver for star formation at these
high redshifts (e.g., \citealt{2008ApJ...673L..21D,
  2010ApJ...713..686D}).  Furthermore, \cite{2011arXiv1105.5395M}
recently used a combination of optical spectroscopy and SED fitting to
measure the host-galaxy properties of luminous (i.e.,
\Lx$\gtrsim10^{44}$~\ergs) Type-2 quasars, reporting remarkably
similar fractions of main-sequence and quiescent hosts as those found
here for moderate luminosity AGNs.  If confirmed, such findings would
imply that the majority of {\it all} nuclear activity has taken place
in galaxies undergoing internally-driven evolution and not in
major-mergers.

\begin{figure}
\begin{center}
	\includegraphics[width=8.2cm]{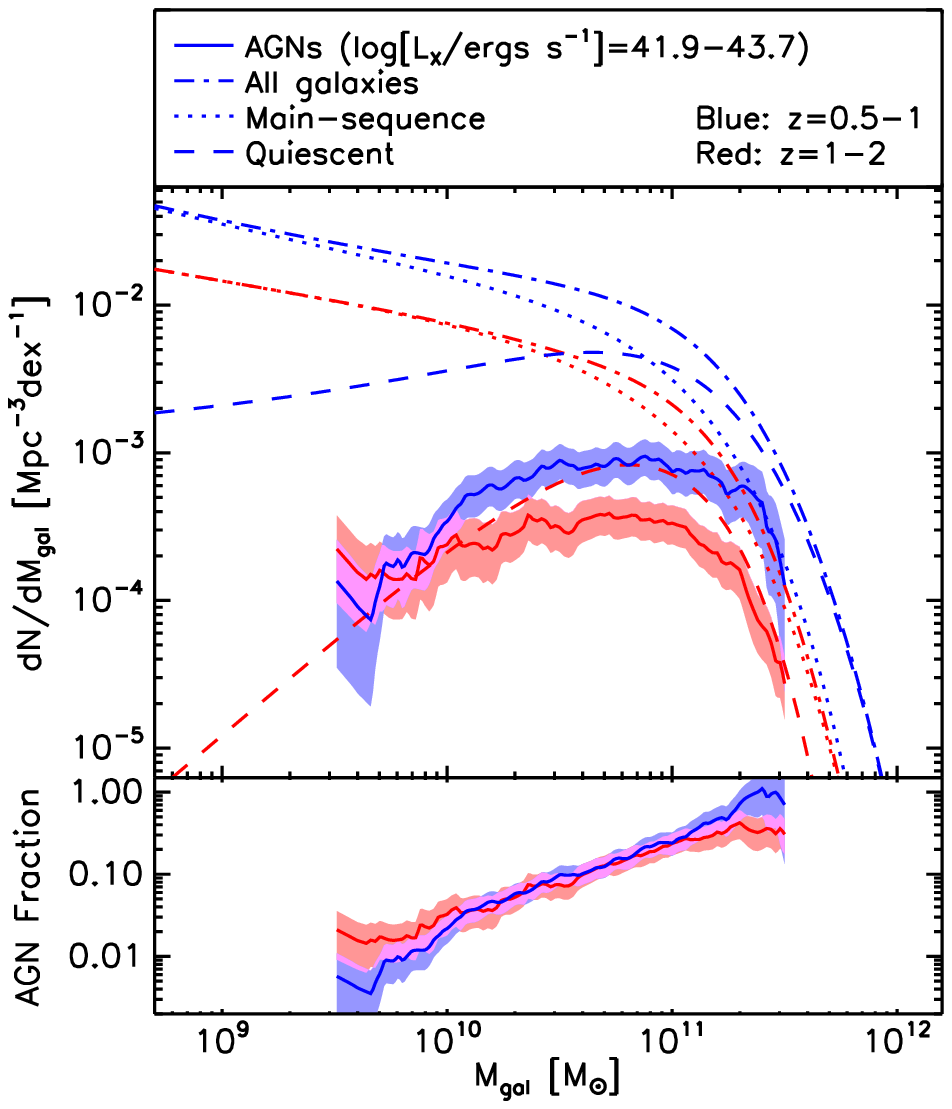}
\end{center}
\caption{{\it Top}: The space density of moderate luminosity AGNs as a
  function of host galaxy stellar mass (i.e., \mgal).  These space
  densities are calculated by multiplying the AGN fractions reported
  in \protect \cite{2010ApJ...720..368X} with the total mass functions
  from \protect \cite{2010ApJ...709..644I}.  The mass functions of
  all, quiescent and star-forming galaxies are also taken from
  \protect \citeauthor{2010ApJ...709..644I}
  (\citeyear{2010ApJ...709..644I}; see key).  Both AGN mass functions
  peak at roughly the point where the total galaxy mass function turns
  over (i.e., $M_\ast$).  {\it Bottom}: The fractions of
  main-sequence, star-forming galaxies that host moderate luminosity
  AGNs as a function of stellar mass.  The fraction of main-sequence
  galaxies hosting moderate luminosity AGNs increases strongly with
  host stellar mass.  This, in turn, implies that the AGN duty cycle
  also increases strongly with host galaxy stellar mass.}
\label{MassFunc}
\end{figure}

\subsection{The fraction of main-sequence galaxies that host AGNs}
\label{Discussion:Mstar}

We have shown that the majority (i.e., $79\pm10$ per cent) of distant,
moderate luminosity AGNs reside in main-sequence galaxies. However,
these AGNs are not evenly distributed among the main-sequence
population. Instead, they preferentially reside in massive galaxies
(i.e., \mgal$\gtrsim10^{10}$~\msun) both locally (e.g.,
\citealt{2003MNRAS.346.1055K}) and at high redshifts (e.g.,
\citealt{2009ApJ...696..396S, 2010ApJ...720..368X}; also our
\fig{Massz}).\footnote{It was recently suggested that stellar mass,
  combined with emission lines, can be used as an effective AGN
  diagnostic out to $z\sim1$ (in the ``Mass-Excitation'' diagram;
  \citealt{2011ApJ...736..104J}).  The lack of evolution in the
  stellar mass of moderate luminosity X-ray AGN since $z\approx3$
  suggests that the diagnostic power of this parameter could be
  extended to higher redshift.}  Therefore, while nuclear activity is
relatively rare in the overall galaxy population ($\sim 6-7$ per cent
of all \mgal$>3\times10^{9}$~\msun\ galaxies), a significant fraction
of high mass galaxies host moderate luminosity AGNs.  We know that
locally, up to 80 per cent of \mgal$=10^{11}$~\msun\ galaxies are
hosting an AGN (\citealt{2005MNRAS.362...25B}).

Recently, a number of studies have shown that the mass function of
main-sequence galaxies has maintained a roughly constant shape since
$z\sim2$, with only its normalisation changing with redshift
(\citealt{2010ApJ...709..644I,2010ApJ...721..193P}).  The range of
characteristic masses at which this mass function turns over (i.e.,
${\rm log[}M_\ast/{\rm M_\odot}]=10.7-11.0$) is remarkably similar to
the typical stellar masses of galaxies hosting moderate luminosity
AGNs (see \fig{Massz}).  Whilst the similarity between these two mass
ranges may be coincidental, it is interesting to explore it further by
considering roughly what fraction of main-sequence galaxies within
this mass range host moderate luminosity AGNs.

In their recent study of the rest-frame colours, magnitudes and host
galaxy masses of AGNs in the \Chandra\ Deep Fields,
\cite{2010ApJ...720..368X} noted that the fraction of galaxies
containing AGNs in similar \lx\ and redshift bins to those considered
here (i.e., $z=0$--1, 1--2, 2--3 and log$[$\lx/\ergs$]=$41.9--43.7)
increases strongly as a function of host galaxy mass, but shows only a
weak change with redshift.  By combining the AGN fractions calculated
in that study with the galaxy stellar mass functions published in
\cite{2010ApJ...709..644I}, we calculate both the mass function and
fraction of galaxies hosting moderate luminosity AGNs as a function of
galaxy stellar mass (see \fig{MassFunc}).\footnote{We use the mass
  function of ``intermediate activity'' galaxies from
  \cite{2010ApJ...709..644I} as they have roughly the same SSFRs as
  main-sequence galaxies at the redshifts considered.} We find that
the fraction of main-sequence galaxies hosting
log$[$\lx/\ergs$]=$41.9--43.7 AGNs increases strongly from $\sim$1--2
per cent at \mgal$\sim$3\e{9}~\msun\ to $\gtrsim$20 per cent at
\mgal$\sim$3\e{11}~\msun, although we note the large uncertainties in
these fractions at high stellar masses due to the relatively small
number of AGNs hosted in \mgal$>10^{11}$~\msun\ galaxies in the CDFs
(see \citealt{2010ApJ...720..368X}, Fig 12k).  Assuming that nuclear
activity is a largely stochastic process, these fractions imply that
the duty cycle of AGNs in more massive galaxies is significantly
longer than in lower mass galaxies.  We can rule out selection effects
as producing this strong increase because (a) there is only a very
weak correlation between \lx\ and host stellar mass (see \Fig{LxM})
and (b) the \cite{2010ApJ...720..368X} sample is largely complete
within the log$[$\lx/\ergs$]=$41.9--43.7 range.  Similarly,
incompleteness due to the effects of X-ray absorption is also unlikely
to produce this effect.  We again stress that these calculations only
apply to AGNs within the log$[$\lx/\ergs$]=$41.9--43.7 luminosity
range.

Whilst we acknowledge that these fractions are only crude estimates,
it is clear that a considerable fraction of main-sequence galaxies
with masses above the turnover of the galaxy mass function host
moderate luminosity AGNs.  Incidentally, this turnover is similar to
the stellar mass range that has dominated the star formation rate
density since $z\sim3$. While we are reluctant to over-interpret this
result, especially in light of the arguments outlined above, if
nuclear activity is necessary to quench the growth of massive
galaxies, then there is no lack of it in these galaxies.  However,
without clear evidence of AGN ``feedback'' from these moderate
luminosity AGNs (e.g., powerful, large scale outflows such as those
reported in \citealt{2006ApJ...650..693N, 2010MNRAS.402.2211A} and
\citealt{2011ApJ...729L..27R}), we are disinclined to take any firm
conclusions on AGN quenching from this analysis.

\section{Summary}
\label{Summary}

We have used the deepest surveys yet undertaken with the \Chandra,
\Spitzer\ and \Herschel\ telescopes to explore the infrared properties
and star formation rates of the host galaxies of X-ray AGNs out to
$z\sim3$.  By using the ultra-deep GOODS-\Herschel\ observations to
identify infrared counterparts to X-ray AGNs in the \Chandra\ Deep
Fields we reach an infrared detection rate roughly 2 times higher than
any previous far-infrared (i.e., FIR) study of AGNs in these fields.
In at least 94 per cent of cases of AGNs at $z>0.5$, the observed
100~\mum\ and 160~\mum\ wavelengths probed by \Herschel-PACS are
dominated by emission from the host galaxy meaning that we can
reliably use \Herschel\ fluxes to measure the (specific) star
formation rates of the host galaxies of AGNs.

The main results from this study can be summarised as follows:

\begin{itemize}
\item There is no clear correlation between AGN power and infrared
  luminosity (and, by inference, star formation rates) for moderate
  luminosity AGNs at all redshifts considered (see
  \S\ref{Results:LIR}).

\item The previously observed increase in the mid to far-infrared
  luminosities with redshift is entirely consistent with being caused
  by the increase in the specific star formation rates (i.e., SSFRs)
  observed in normal, star-forming (i.e., main-sequence) galaxies (see
  \S\ref{Results:Masses}).

\item The majority (i.e $79\pm10$ per cent) of moderate luminosity
  (i.e., \lx$=10^{42}$--$10^{44}$~\ergs) AGNs reside in main-sequence
  galaxies.  We estimate that only $7\pm1$ per cent reside in
  starbursting galaxies and $15\pm7$ per cent reside in quiescent
  galaxies.  These fractions display a slight preference toward
  main-sequence compared to the general galaxy population and a
  remarkably similar to those recently reported for Type 2 Quasars
  (see \S\ref{Results:Masses} and \S\ref{Discussion:Hosts}).

\item The average SSFRs of moderate AGN hosts are only marginally
  lower (i.e., $\approx20$ per cent lower) than those of normal
  main-sequence galaxies, with this small deficit being due to a
  fraction of AGNs residing in quiescent galaxies.  While this may
  hint at a causal connection between moderate levels of nuclear
  activity and star formation (e.g., AGN quenching ongoing star
  formation), the impact on SSFRs is clearly very limited; (see
  \S\ref{Results:Masses} and \ref{Discussion:Impact}).

\item The observed 16--160~\mum\ SEDs are also broadly consistent with
  being more strongly dominated by the host galaxy at higher redshifts
  (see \S\ref{Results:SEDs}).  Parenthetically, we note that this
  means that using infrared diagnostics to identify AGNs will be
  increasingly difficult at higher redshift.

\item The host galaxies of moderate luminosity AGNs have the same red
  rest-frame colours as non-AGN-hosting galaxies of similar mass.
  Despite actively forming stars, these massive (i.e.,
  $\gtrsim10^{10}$~\msun) galaxies have red colours, likely due to
  their high concentrations of dust (see \S\ref{Results:RedColours}).

\item The fraction of main-sequence galaxies experiencing moderate
  levels of nuclear activity increases strongly with host galaxy
  stellar mass, from a few per cent at \mgal$\sim3\times10^{9}$~\msun\
  to $>50$ per cent at \mgal$\sim3\times10^{11}$~\msun (see
  \S\ref{Discussion:Mstar}).

\end{itemize}

\noindent

In our discussion we argue that the results presented here strongly
suggest that the majority of moderate nuclear activity in the Universe
has taken place in normal star forming galaxies undergoing internal
evolution, rather than in violent mergers (see
\S\ref{Discussion:Hosts}).  Since merger-driven inflows are ruled out
in the majority of cases, we speculate that disk instabilities play an
important role in fuelling nuclear activity, as suggested by some
theoretical models.  What is remarkable is that we do not see any
correlation between the levels of AGN and star formation activity
taking place within these galaxies.  Indeed, the only significantly
different aspect of AGN host galaxies compared to the general galaxy
population are their high stellar masses, which appears to have been
the main distinguishing feature of galaxies hosting moderate AGNs
since $z\sim3$.

\section*{acknowledgements}
We are grateful to N. Drory for sharing the SED-fitting code used to
estimate galaxy stellar masses.  We thank V. Strazzullo and the
anonymous referee for their useful comments.  We acknowledge funding
from a European FP7 Co-fund Fellowship (JRM) and the Science and
Technology Funding Council (DMA). DE and MP acknowledge financial
support from the French Agence Nationale de la Recherche (ANR) project
{\it HUGE}, ANR-09-BLAN-0224. This research was supported by the
ERC-StG grant UPGAL 240039 and by the French ANR under contract
ANR-08-JCJC-0008. This work is based on observations made with {\it
  Herschel}, a European Space Agency Cornerstone Mission with
significant participation by NASA. Support for this work was provided
by NASA through an award issued by JPL/Caltech.

\begin{table*}
  \caption{The average AGN and host galaxy properties of X-ray
    sources in the \Chandra\ Deep Fields, separated into 
    the various classification, \lx\ and redshift bins used throughout this study.}
  \begin{center}
    \begin{tabular}{@{}clccccccccc@{}}
\hline
\hline
&&&&&&(7)&(8)&(9)&(10)&(11)\\
(1)&(2)&(3)&(4)&(5)&(6)&$L_{\rm 2-10keV}$&$M_{\rm gal}$&$L_{\rm IR}$&SFR&SSFR\\
Index&Description&$N_{\rm All}$&$N_{\rm Und}$&$N_{\rm Mass}$&$z$&$(10^{42}{\rm~ergs~s^{-1}~cm^{-2}})$&$(10^{10}{\rm~M_{\odot}})$&$(10^{10}{\rm~L_{\odot}})$&$({\rm~yr^{-1}})$&$({\rm~Gyr^{-1}})$\\
\hline
 1&All                       &         504&         276 (55)&         338 (67)& 1.41& 1.98&$6.87^{+0.72}_{-0.65}$&$58.0\pm3.4$&$100.1\pm5.9$&$1.46^{+0.18}_{-0.16}$\\
 2&AGN                       &         367&         228 (62)&         255 (69)& 1.72& 7.04&$7.74^{+0.96}_{-0.85}$&$55.2\pm3.8$&$95.4\pm6.6$&$1.23^{+0.18}_{-0.16}$\\
 3&Starbursts                &          41&          10 (24)&          26 (63)& 0.67& 0.11&$4.9^{+2.5}_{-1.7}$&$19.1\pm3.3$&$33.1^{+5.7}_{-5.8}$&$0.67^{+0.38}_{-0.25}$\\
 4&$z=0.5-3$ $L_X=10^{42-44}$&         219&         127 (58)&         159 (73)& 1.61& 8.66&$8.6^{+1.2}_{-1.0}$&$49.7^{+5.7}_{-5.3}$&$85.9^{+9.8}_{-9.1}$&$1.00^{+0.18}_{-0.16}$\\
 5&$z=0.5-1$ $L_X=10^{42-44}$&          50&          20 (40)&          46 (92)& 0.81& 4.93&$7.8^{+2.1}_{-1.7}$&$12.3\pm2.2$&$21.2^{+3.8}_{-3.7}$&$0.271^{+0.091}_{-0.071}$\\
 6&$z=1-2$ $L_X=10^{42-44}$  &         103&          60 (58)&          83 (81)& 1.45& 7.50&$8.6^{+1.6}_{-1.3}$&$42.3^{+5.9}_{-6.4}$&$73^{+10}_{-11}$&$0.85^{+0.20}_{-0.18}$\\
 7&$z=2-3$ $L_X=10^{42-44}$  &          66&          47 (71)&          30 (45)& 2.49&16.59&$11.1^{+3.5}_{-2.6}$&$136\pm27$&$236^{+46}_{-47}$&$2.12^{+0.82}_{-0.63}$\\
 8&$z=0.5-1$ $L_X=10^{42-43}$&          36&          18 (50)&          35 (97)& 0.81& 2.86&$7.2^{+2.6}_{-1.9}$&$11.9\pm2.7$&$20.5\pm4.6$&$0.29^{+0.13}_{-0.10}$\\
 9&$z=1-2$ $L_X=10^{42-43}$  &          61&          39 (64)&          48 (79)& 1.44& 3.09&$8.7^{+2.1}_{-1.7}$&$41.7^{+7.6}_{-8.3}$&$72^{+13}_{-14}$&$0.83^{+0.26}_{-0.22}$\\
10&$z=2-3$ $L_X=10^{42-43}$  &          22&          12 (55)&          10 (45)& 2.46& 3.85&$8.0^{+3.6}_{-2.5}$&$122^{+41}_{-45}$&$211^{+71}_{-77}$&$2.6^{+1.6}_{-1.2}$\\
11&$z=0.5-1$ $L_X=10^{43-44}$&          14&           2 (14)&          11 (79)& 0.80&20.00&$9.9^{+4.5}_{-3.1}$&$13.0\pm2.9$&$22.4^{+5.1}_{-5.0}$&$0.23^{+0.12}_{-0.08}$\\
12&$z=1-2$ $L_X=10^{43-44}$  &          42&          21 (50)&          35 (83)& 1.46&27.14&$8.2^{+2.6}_{-2.0}$&$45.4^{+8.2}_{-8.5}$&$78^{+14}_{-15}$&$0.95^{+0.36}_{-0.28}$\\
13&$z=2-3$ $L_X=10^{43-44}$  &          44&          35 (80)&          20 (45)& 2.50&34.44&$11.6^{+5.0}_{-3.5}$&$147^{+39}_{-38}$&$253^{+67}_{-65}$&$2.2^{+1.2}_{-0.8}$\\
14&$z=0.5-1$ $L_X=10^{41-46}$&          91&          45 (49)&          77 (85)& 0.76& 1.78&$7.5^{+1.6}_{-1.3}$&$8.6^{+1.1}_{-0.8}$&$14.9^{+2.0}_{-1.5}$&$0.198^{+0.050}_{-0.038}$\\
15&$z=1-2$ $L_X=10^{41-46}$  &         134&          76 (57)&         111 (83)& 1.42& 5.29&$8.2^{+1.4}_{-1.2}$&$43.6^{+6.3}_{-6.6}$&$75\pm11$&$0.92^{+0.21}_{-0.19}$\\
16&$z=2-3$ $L_X=10^{41-46}$  &          81&          61 (75)&          38 (47)& 2.52&26.15&$10.7^{+2.2}_{-1.9}$&$141\pm23$&$244\pm39$&$2.29^{+0.63}_{-0.52}$\\
\hline
\end{tabular}
 
  \end{center}{\sc Notes}:(1) Index, (2) Description of bin; ``All''
  includes all X-ray detected sources, irrespective of classification
  (i.e., including AGNs, galaxies and stars), rows 4-16 only include
  AGNs (3) Number of X-ray AGNs with known redshifts covered by the
  GOODS-H survey, excluding those that show signs of being
  AGN-dominated at FIR-wavelengths (see \S\ref{Calculating:LIR}) (4)
  As for column 3, but only counting those that are undetected at FIR
  wavelengths (i.e., the number of sources in each stack).  The
  numbers in brackets are the fractions $N_{\rm Und}/N_{\rm All}$ (in
  per cent) (5) As for column 3, but only counting those for which we
  obtain a ``secure'' host galaxy stellar masses (see
  \S\ref{Calculating:Masses}). The numbers in brackets are the
  fractions $N_{\rm Mass}/N_{\rm All}$ (in per cent) (6) Mean redshift
  (7) Mean rest-frame, absorption corrected 2-10~keV luminosity (8)
  Mean stellar mass (9) Mean infrared luminosity calculated from the
  stacked 100~\mum\ or 160~\mum\ (depending on redshift, see
  \S\ref{Calculating:LIR}) of the undetected sources combined with the
  mean 100~\mum\ or 160~\mum\ flux of the detected sources (10) Host
  galaxy star formation rates derived from the infrared luminosities
  in column 10 using the prescription of \cite{1998ARA&A..36..189K}
  (11) Specific star formation rates calculated by dividing the star
  formation rates in column 11 by the stellar masses in column 9.
  \label{TableAvg} 
\end{table*}

\bsp

\label{lastpage}

\end{document}